\documentclass[11pt,a4paper]{article}
\usepackage{jcappub}
\usepackage{capt-of}
\usepackage{ragged2e}
\usepackage{color}
\usepackage[english]{babel} 
\usepackage{amsmath}
\usepackage{float}
\usepackage{amssymb}
\usepackage{braket}
\usepackage{amsfonts,bbm}
\usepackage{dsfont,tensor}
\usepackage{enumerate}
\makeatletter\AtBeginDocument{\let\@elt\relax}\makeatother
\usepackage{graphicx}
\usepackage{caption}
\usepackage{adjustbox}
\usepackage{bbm}
\usepackage{array}
\usepackage{booktabs}
\usepackage{units}
\usepackage{textcomp}
\usepackage{mathtools}
\usepackage{cancel}
\usepackage{slashed}
\usepackage{multirow}
\usepackage{slashed}
\usepackage{comment}
\usepackage{relsize}
\usepackage{tcolorbox}
\usepackage{csquotes}
\usepackage{simpler-wick}
\usepackage{subcaption}
\usepackage{lipsum}
\usepackage{tabularx}
\allowdisplaybreaks

\definecolor{myRed}{RGB}{205, 126, 143}
\definecolor{myGreen}{RGB}{138, 185, 155}
\definecolor{myGray}{RGB}{128, 128, 128}
\definecolor{myViolet}{RGB}{164, 140, 177}
\definecolor{amber}{rgb}{1.0, 0.75, 0.0}

\newcommand{\pts}[1]{\phantom{.}\hfill(\textit{#1~point}\ifthenelse{\equal{#1}{1}}{}{\textit{s}})}

\newcommand{\mdm}{m}

\newcommand{\Msun}{{\ifmmode{{\rm{M_{\odot}}}}\else{${\rm{M_{\odot}}}$}\fi}}

\newcommand{\beq}{\begin{equation}}
\newcommand{\eeq}{\end{equation}}
\newcommand{\bea}{\begin{eqnarray}}
\newcommand{\ena}{\end{eqnarray}}

\newcommand{\lsim}{\mathrel{\mathop{\kern 0pt \rlap
{\raise.2ex\hbox{$<$}}}
\lower.9ex\hbox{\kern-.190em $\sim$}}}
\newcommand{\gsim}{\mathrel{\mathop{\kern 0pt \rlap
{\raise.2ex\hbox{$>$}}}
\lower.9ex\hbox{\kern-.190em $\sim$}}}

\title{Compact Stars as Portals to Extra-Dimensional Dark Matter}
\author[a]{Raghuveer Garani,}
\author[b]{Chris Kouvaris,}
\author[c]{Michel~H.G.~Tytgat}
\author[d]{and J\'{e}r\^{o}me~Vandecasteele}

\affiliation[a]{Centre for Strings, Gravitation and Cosmology, Department of Physics, Indian Institute of Technology Madras, Chennai 600036, India}
\affiliation[b]{Physics Division, National Technical University of Athens, 15780 Zografou Campus, Athens, Greece}
\affiliation[c]{Service de Physique Th\'eorique, Universit\'e Libre de Bruxelles, Boulevard du Triomphe, CP225, 1050 Brussels, Belgium}
\affiliation[d]{Fakultät für Physik, Universität Bielefeld, 33501 Bielefeld, Germany}

\emailAdd{garani@iitm.ac.in}
\emailAdd{kouvaris@mail.ntua.gr}
\emailAdd{michel.tytgat@ulb.be}
\emailAdd{jvandecasteele@physik.uni-bielefeld.de}

\abstract{We investigate hydrostatic configurations of asymmetric dark matter spheres in scenarios where fermionic dark matter particles can propagate into extra spatial dimensions, while Standard Model fields remain confined to ordinary three dimensions. As the number of extra dimensions increases, the effective equation of state for non-relativistic matter softens, making even modest dark matter accumulation inside neutron stars susceptible to gravitational collapse into extra-dimensional black holes. These black holes are longer lived than their standard $3$ dimensional counterparts and can accrete enough material to consume an entire neutron star, ultimately producing solar-mass scale black holes. For geometric cross sections, dark matter with masses above $\mathcal{O}(10\,{\rm TeV})$ may already be excluded for more than two extra dimensions of size ${\mathcal{O}(\rm fm})$---sharply contrasting with the  $3$ dimensional case, where comparable limits only appear for masses $\gtrsim 10^{5}$ TeV at typical halo densities of $0.3\, \rm{GeV/cm^3}$.}

\begin{document}
\maketitle

\section{Introduction}
\begin{center}
\begin{figure*}[htb!]
    \centering
    \includegraphics[width=1.0\linewidth]{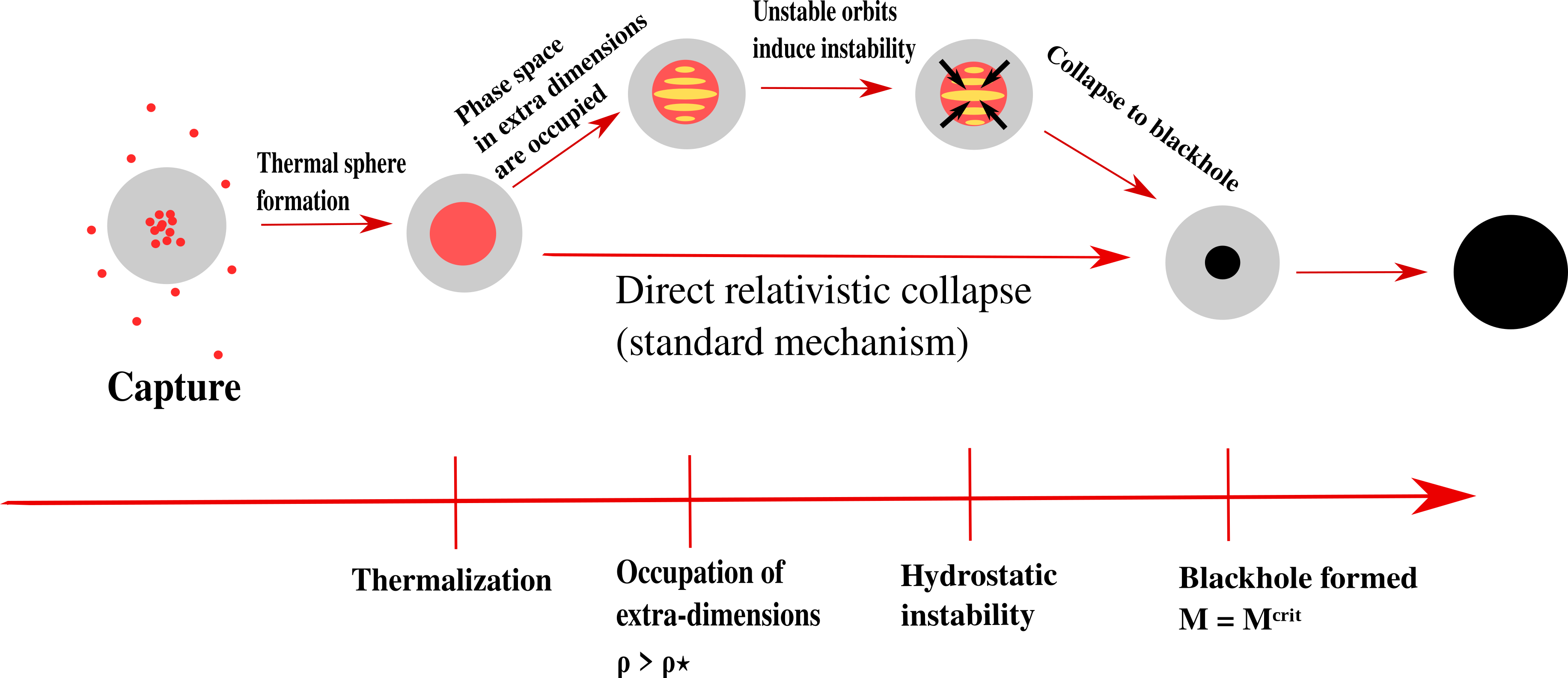}
    \caption{\justifying Evolution of a dark matter sphere in NSs as the extra-dimensional phase space becomes accessible. After thermalization, DM particles become degenerate; when their density exceeds $\rho_\star$ (eq.~\eqref{eq:rhostar}) the XD are populated. Further accumulation of particles triggers instability and collapse to a BH of mass $M_{\rm crit}$, given by eq.~\eqref{eq:Mcrit_sg}.
    }
    \label{fig:draw}
\end{figure*}
\end{center}

The mere existence of old neutron stars (NS) can exclude asymmetric  dark matter (DM) particle candidates~\cite{Goldman:1989nd,Gould:1989gw,Kouvaris:2010jy,Kouvaris:2011fi, Kouvaris:2011gb,McDermott:2011jp, Bell:2013xk, Bramante:2013hn, Bramante:2013nma, Bramante:2017ulk, Garani:2018kkd, Dasgupta:2020mqg,Tinyakov:2021lnt,Garani:2021gvc,Bramante:2023djs,Bhattacharya:2023stq,Bramante:2024idl,Robles:2025dlv}. The argument is standard and is as follows (see figure~\ref{fig:draw}): once DM particles are captured from the galactic halo, they thermalize with the NS medium and form an isothermal core~\cite{Bertoni:2013bsa,Garani:2020wge, Bell:2023ysh}. As DM particles are continuously accreted, the density of the thermal sphere increases and, if the density exceeds the neutron start central density, it will be unstable and collapse to a black hole (BH). Even below this threshold, if the Chandrasekhar condition for the number of DM particles is met, the sphere can form a BH~\cite{Garani:2021gvc}. 

An issue with this scenario is that, unless the DM is extremely heavy,  the number of DM particles required to trigger black hole formation is typically much larger than what can be accumulated under standard astrophysical assumptions. For DM densities of $\sim 0.3$ GeV/cm$^3$, only fermions with $m \gtrsim 10^{5}$ TeV can be excluded for geometric DM-nucleon cross section. However, such heavy asymmetric DM is disfavored because the large annihilation cross sections required to deplete the symmetric component of the DM is  incompatible with unitarity bounds, typically $m \lesssim 10^2$ GeV \cite{Griest:1989wd}. It is therefore  of interest, both phenomenologically and as matter of principle, to explore whether, in one way or another, sensitivity to lighter asymmetric DM particles could be achieved. 

One way to put it is that one needs to circumvent the Chandrasekhar criterion for collapse of degenerate fermions into a BH. This  arises when the Fermi pressure cannot support gravitational attraction. For standard scenarios, this occurs only when particles become relativistic due to a softening of the ferminonic matter equation of state. For a fermion of mass $m$, the critical Chandrasekhar mass in 3 ordinary dimensions (3D) is $M_{\rm ch} \sim M^3_{\rm Pl}/m^2$ (for $m \sim $ GeV, $M_{\rm ch} \sim M_\odot$). There have been some attempts to alter the parametric form of this Chandrasekhar limit. 
These introduce a new attractive short-range interaction that, in principle, could hasten collapse~\cite{Kouvaris:2011gb, Kouvaris:2015rea,Kouvaris:2018wnh}. 
The simplest possibility is to assume that this  interaction is induced by the exchange of a spin 0 or scalar particle with a Yukawa coupling to DM fermions. However, as shown in \cite{Gresham:2018rqo}, such a Yukawa interaction eventually causes the DM mass to decrease, eventually rendering the DM relativistic, at which point the Fermi pressure overcomes the attractive interaction; the outcome is that the Chandrasekhar (see also \cite{Garani:2022quc}). 

Since spin-1 exchange between same charge particles is repulsive, the next logical possibility is to consider the exchange of a spin-2 particle. In this work, we take a different route. Instead of considering the exchange of a single massive spin-2 particle, we consider a full Kaluza-Klein tower of such states. Freely inspired by brane-world~\cite{Arkani-Hamed:1998jmv,Antoniadis:1998ig,Randall:1999ee,Randall:1999vf} and dark dimensions frameworks~\cite{Montero:2022prj}, we consider a scenario in which fermionic DM particles propagate in $d$ large extra dimensions (XD), while Standard Model particles remain confined to 3 spatial dimensions. While $d=1$
is phenomenologically excluded, $d\geq 2$ with characteristic scale $R_\star \lesssim \mu m$ is allowed~\cite{Yagi:2011yu}. Three simple effects motivate this setup. They all point towards the possibility of gravitational collapse for lighter DM particles. First and foremost, in the Kepler problem, no stable bound orbits exist for $d \geq 2$~\cite{Misner:1973prb,Tangherlini:1963bw,ehrenfest1917way}. As was already known to Newton, even an inverse-cube force law -- corresponding to $d=1$ extra dimension -- can overcome the centrifugal barrier in the 2-body problem, leading to orbital instability \cite{Newton:1687eqk}. Second, if there are large extra dimensions, the higher-dimensional Planck scale can be significantly lower than its 3D counterpart, enhancing the strength of gravitational interactions~\cite{Arkani-Hamed:1998jmv,Antoniadis:1998ig}. This, in principle, could even lead to the formation of black holes at high energy particles colliders \cite{Argyres:1998qn,Dimopoulos:2001hw}.  Last but not least, the equation of state of degenerate fermions becomes softer in higher dimensions than in 3D. Together, these effects suggest that a sufficiently dense DM distribution, if it could probe extra dimensions, could undergo gravitational instability, leading to black hole formation. 

The extent to which such a scenario can modify the standard Chandrasekhar criterion, and thereby the constrained parameter space of asymmetric fermionic DM, remains to be determined.
The situation we propose to explore is simple and is as depicted in Fig.~\ref{fig:draw}. It begins with the standard capture of DM particles by a neutron star, forming a 3D spherical DM cloud. We consider standard astrophysical conditions for the local DM density and capture mechanism. As DM accumulates, the core of the cloud may reach a critical density $\rho_\star$ (defined below), at which point DM particles begin to populate the extra-dimensional phase space, leading to a softening of the equation of state. As we will show, this is the more tractable part of the problem. Simultaneously, the DM cloud also departs from its 3D spherical distribution in configuration space, and particles in its core also begin to feel the extra-dimensional gravitational interaction\footnote{We deliberately neglect interactions other than gravity along the extra dimensions.}. Due to the lack of symmetry, this aspect of the problem is more difficult to address analytically, and is work in progress. In this work, instead, we focus on the modification of the equation of state alone, and show that for $d \geq 3$, the softening induced by the filling of the extra-dimensional phase space is sufficient to guarantee gravitational collapse of the DM cloud to a black hole at a lower DM mass than in 3D. Given the inherent complexity of the complete higher-dimensional problem, it is noteworthy that such a robust result can be established.\\


The plan of our work is a follows. We set the ground in section \ref{sec:Xdims}, in which we write down and determine the higher-dimensional equation of state (see Fig.~\ref{fig:polytrop_d} and also Appendix \ref{app:xdims0}).
In section \ref{sec:hydro}, we track the hydrostatic evolution of  a DM 3D spherical distribution and argue that even a relatively small occupation number in extra dimensions can lead to gravitational instability, even for non-relativistic (NR) particles, likely culminating with the formation of a higher-dimensional BH.  We  consider two scenarios. In scenario (I), the DM sphere self-gravitates first and then probes extra dimensional phase space. In scenario (II), extra dimensional phase space is occupied before the onset of self-gravitation. For both cases, we solve the Lane-Emden equation for the DM sphere and perform a linear stability analysis. We find that the averaged adiabatic index $\langle \Gamma \rangle$ becomes $\leq 4/3$ for $d\geq 3$, indicating an instability (see Section 6.7 and Eq.~(6.7.10) in \cite{book:shapiro}), leading to collapse to  BH in the self-gravitation scenario, see Fig.~\ref{fig:M-R_LM_d} where we depict the Mass-Radius relation for the DM clouds for various choice of $d$, including $d=0$ (3D scenario). Further aspects of the Chandrasekhar criterion in our extra-dimensional setup are discussed in Appendix \ref{app:LM_sols}. An extra dimensional BH, once formed and trapped in a NS, would live longer than BH of the same mass in 3D. In effect, we show in section \ref{sec:BH_stuff} that it would have sufficient time to consume all of the NS material, leading in turn to the formation of solar-mass BH~\cite{Kouvaris:2018wnh}, see figure~\ref{fig:draw}. As a consequence, we show in section \ref{sec:constraints} that asymmetric fermion DM of mass $\sim 10\,{\rm TeV}$ or higher are constrained, depending on assumptions on the scale and number of extra dimensional. This contrasts with the standard 3D case, in which constraints apply only to extremely heavy fermionic DM. Reversing the logic, a glimpse at Fig.~\ref{fig:CollapseParamSpace} reveals that DM captured in neutron stars probes extra dimensions significantly smaller than those accessible to standard experiments such as the LHC or Eötvos-type tests. As such, compact stars may open a portal to extra-dimensional dark matter.\\

 \section{Extra-dimensional equation of state }
 \label{sec:Xdims}

 \subsection{Excursions in extra dimensions}
 We consider 
$d$ extra compact dimensions, toroidally compactified as 
$T^d$ with radius $R_\star = R_d/2\pi$. 
The momentum components along $T^d$ would be $p_i =  n_i/R_\star$, with $i=1,\ldots d$ and $n_i \in \mathbb{Z}$. At low densities, fermions populate only the zero modes ($n_i =0$),  therefore the system is effectively 3D. As the density rises, the Fermi momentum increases until the Fermi energy exceeds the first excited mode and occupation of higher dimensional momentum states becomes possible. For a NR particle with $p = (\vec p, p_\perp)$, where $p_\perp = (p_1,\ldots,p_d)$, the energy is $E = p^2/(2m) + \sum_{i=1}^d (n_i/R_\star)^2/(2m)$, so the excitation energy is 
\begin{equation}\label{eq:EKKmode1}
  E_{\star} = \frac{1}{2 m}\left(\frac{1}{R_\star}\right)^2~.
  \end{equation}
For a 3D NR degenerate Fermi gas, the Fermi energy is 
\begin{equation}\label{eq:Ef3d}
  E_F = \frac{p^2_F}{2 m} = \frac{1}{2 m} \left(\frac{3 \pi^2}{g} n \right)^{2/3}~,
 \end{equation}
with $n=N/V$ the number density and $g$ the degeneracy factor ($g=2$ in the sequel). Fermions begin to occupy higher levels when $E_F \gtrsim E_\star$. Consequently, the energy density $\rho_\star = m_{} n_\star$ at which XD phase space starts to be occupied is 
\begin{equation}\label{eq:rhostar}
  \rho_\star= \frac{g\,m}{3 \,\pi^2R^3_\star} \approx  10^{8} \left(\frac{m}{1\, {\rm TeV}} \right) \left(\frac{\rm 10^3 \, fm}{R_\star} \right)^3\, {\rm g/cm^3}~.
\end{equation}

After capture in a NS, the DM particles thermalize with the neutrons with  temperature $T_{\rm NS}$ and form a thermal sphere with scale radius $r_{th}\sim (m G \rho_{\rm NS}/T_{\rm NS})^{-1/2}$. The  average mass density is $\rho_{\rm th} = m C\times t/V_{th}$, with $C$ the DM accretion rate.
The DM gas becomes degenerate when the thermal de Broglie wavelength $\lambda_{\rm dB}=(m T/2\pi)^{-1/2}$ reaches $\gtrsim n^{-1/3}$. In the NR limit, the energy density required is $\rho_{\rm deg} \gtrsim m/\lambda^3_{\rm dB}$. 
After the onset of degeneracy, gravitational forces are countered by the Fermi pressure, giving a typical scale for the degenerate sphere, $r_F \sim (G m^2 \rho_{\rm NS})^{-1/4} N^{1/6}$. Here $N= C\times t$ is the total number of DM particles accreted at time $t$. Therefore, the energy density
 ($\sim M/r_F^3$) for $T \sim 0$ is given by 
\begin{equation}
    \rho_{\rm deg} \approx 2\times 10^{11} \left(\frac{m}{100\,\rm GeV}\right)^{5/2} \left(\frac{N}{ 10^{40}}\right)^{1/2}  {\rm g/cm^3}~.
\end{equation}
These relations show that, for a wide range of DM masses and XD sizes, fermionic DM occupy extra-dimensional phase space once degenerate.\footnote{For large XD sizes, the DM cloud may not reach degeneracy at $\rho_\star$; see section~\ref{app:xdims0} for details.}\\

\subsection{EoS in degenerate limit}
 As we now show, the equation of state of a degenerate Fermi gas in ${\rm I\!R}^3 \otimes T^d $ is polytropic with equation of state (EoS) $P = K_{d} \rho^{\gamma_d}$; both the polytropic (or adiabatic) index $\gamma$ and $K$ depend on $d$; for ${\rm I\!R}^{3+d}$, $\gamma_d = (5+d)/(3+d)$. On dimensional grounds the effective 3D pressure is expected to be of the form 
 \begin{equation}\label{eq:Pgen}
  P_3 = K_{d}\,\rho^{\gamma_d}_{3} \times V^{1-\gamma_d}_{\rm extra} \quad \quad {\rm with} \quad  \quad    V_{\rm extra} = (2\pi R_\star)^d~.
 \end{equation}
 The constant ($K_{d}$) is defined as
\begin{eqnarray}\label{eq:kconst}
K_{d}&=&\frac{2\pi}{2^{\frac{2}{3+d}}}\frac{\left(\Gamma\left(\frac{5+d}{2}\right)\right)^{\gamma_d}}{\Gamma\left(\frac{7+d}{2}\right)}\frac{1}{m^{\gamma_{d}+1}}\quad \quad {\rm with} \quad  \quad \gamma_d=\frac{5+d}{3+d}~.
\end{eqnarray}
  
\begin{figure*}[htb!]
      \centering
      \includegraphics[width=1.0\linewidth]{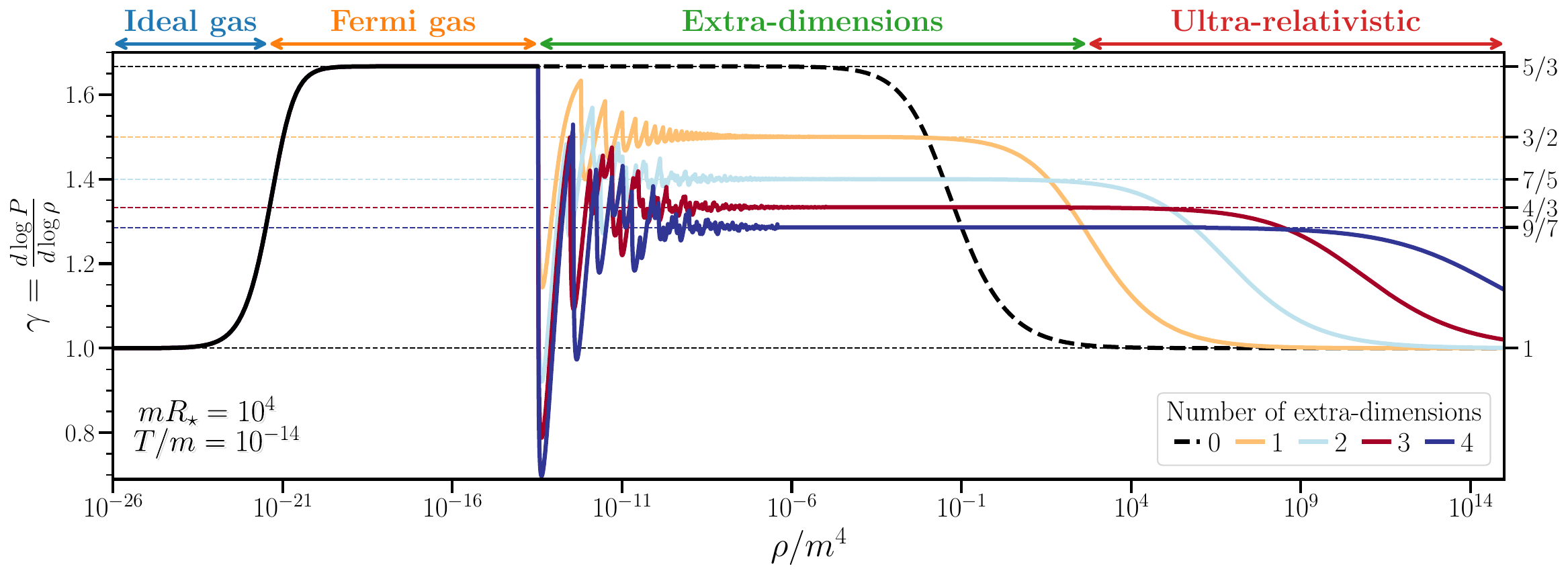}              \caption{\justifying Adiabatic index as a function of density for various cases of extra spatial dimensions for $m R_\star = 10^4$.
      }
        \label{fig:polytrop_d}
\end{figure*}

 In our XD setup, the energy density and pressure (both along 3D) of a Fermi gas with $\beta =1/T$ and chemical potential $\mu$ are given by
\begin{align}
     \rho &=  g \sum_{p_\perp} \int \frac{d^3 p}{(2 \pi)^3} \frac{E}{e^{\beta(E-\mu)} +1}~, \label{eq:den3d}\\
     P &= g \sum_{p_\perp} \int \frac{d^3 p}{(2 \pi)^3} \frac{p^2}{3 E}\frac{1}{e^{\beta(E-\mu)} +1}~,\label{eq:pres3d}
\end{align}
where $E -\mu \approx p^2/(2m) + p^2_\perp/(2m) -\bar{\mu}$ in the NR limit, with $\bar \mu = \mu - m$.
We have evaluated the  adiabatic index, $\gamma = d \log P/d\log\rho$ as function of $\rho$, by evaluating eqs.~\eqref{eq:den3d} and~\eqref{eq:pres3d} as function of $\mu$ for a given  $m R_\star$, see fig.~(\ref{fig:polytrop_d}). 

At low densities, $P \sim \rho T/m$ and the polytropic index is $\gamma = 1$. As the density increases, for fixed $T$ and 3D volume, degeneracy sets in and $\gamma$ approaches the 3D NR Fermi gas value, $\gamma_0 = 5/3$. When $\rho \gtrsim \rho_\star$, modes along the XD begin to be filled. The larger available phase space softens the polytropic index ($\gamma < \gamma_0$), with an oscillatory pattern due to the discreteness of states. At even higher densities, the continuum limit is reached and $\gamma \rightarrow \gamma_d$. Finally, for $\rho \gg m^4$, particles become relativistic and $P = \rho/\left(3+d\right)$. General expressions for all thermodynamic quantities, including their finite-temperature behavior, are given in appendix~\ref{app:xdims0}.

\section{Hydrostatic evolution}
\label{sec:hydro}
 Hydrostatic equations relate the pressure (kinetic part) to the potential energy, given by, in 3D, 
\begin{equation}\label{eq:hydro_gen}
    \vec{\nabla} P + \rho \vec{\nabla} \Phi = 0~\quad \quad {\rm and }\quad\quad \nabla^2 \Phi = 4 \pi G \rho~.
\end{equation}
In the standard 3D  scenario, all short distance interactions (e.g. local interactions) are encoded in the EoS while interactions between particles that are separated  over long distances (i.e. gravity) are encoded in the potential through the Poisson equation. In the XD scenario, in addition to the EoS, the effective long distance correction to Newton's constant $G$ due to extra-dimensions should also be taken into account, see also~\cite{Kan:2002yi}.

Current experimental constraints allow the Newtonian gravitational potential to be modified at relatively large distances $R_\star \sim \mu m$ for $d \geq 2$~\cite{Adelberger:2006dh, Yagi:2011yu}. For $r \gtrsim R_\star$, the modification of the gravitational interaction can be modeled through a Yukawa potential, with range $R_\star$. However, for even smaller particle separations $r\ll R_\star$, the gravitational potential should scale as $1/r^{1+d}$.  Generally,
\begin{align}
\Phi\left(\vec{r}\right)&=\int d^dr' \rho\left(\vec{r}'\right) V\left(\vec{r}-\vec{r'}\right)
\end{align}
where $d$ is the number of large spatial dimensions. Here, $V$ is the potential generated by one particle. In principle, such considerations should be included in the hydrostatic equations~\eqref{eq:hydro_gen} at extremely small distances and matched to the usual 3D hydrostatic equations at large distances, in order to precisely capture the evolution of degenerate matter in an XD scenario. This turns out to be technically complicated.

In light of the drastic softening of the polytropic index $\gamma$ (for $d \geq 3$), we argue that  modifications of the EoS alone may be sufficient to draw conclusive statements about hydrostatic stability. As illustrated in Fig.~\ref{fig:polytrop_d},
the adiabatic index is $\gamma \leq 4/3$ for $d\geq3$. It is a known fact that such polytropes are unstable to perturbations~\cite{book:shapiro}.  Due to the attractive nature of gravitation, we expect that solving the hydrostatic equations including the very short distance gravitational potential in XD will only make the instability more apparent~\cite{Feng:2022fuk}. 
Therefore, as a first step in XD scenarios, in this work we  consider the standard 3D hydrostatic equations without including any corrections to the gravitational potential. As we will show, some robust statements are already possible for scenarios with $d \geq 3$.  A more general discussion, also addressing the intermediate cases  $d=1$ and $2$, will be given in a future work~\cite{GKTV:2026}.\\

\subsection{Mass-radius relation and instability}
We define the following dimensionless quantities: $\bar{r}=r/R_L$, $R_L=M_{\rm Pl}/m^2$ and $\bar{M}=M/M_L$, $M_L=M_{\rm Pl}^3/m^2$, and $c_s^2=d\bar{P}/d\bar{\rho}$, $\bar{P}=P/m^4$, and $\bar{\rho}=\rho/m^4$. The standard Chandrasekhar limit for degenerate fermions corresponds to $M_{\rm Ch} = M_L$.  We consider two possibilities, depending whether the DM cloud is self-gravitating or is dominated the gravitational potential of the background neutrons. \\

\begin{figure}[htb]
      \centering
          \includegraphics[width=0.6\linewidth]{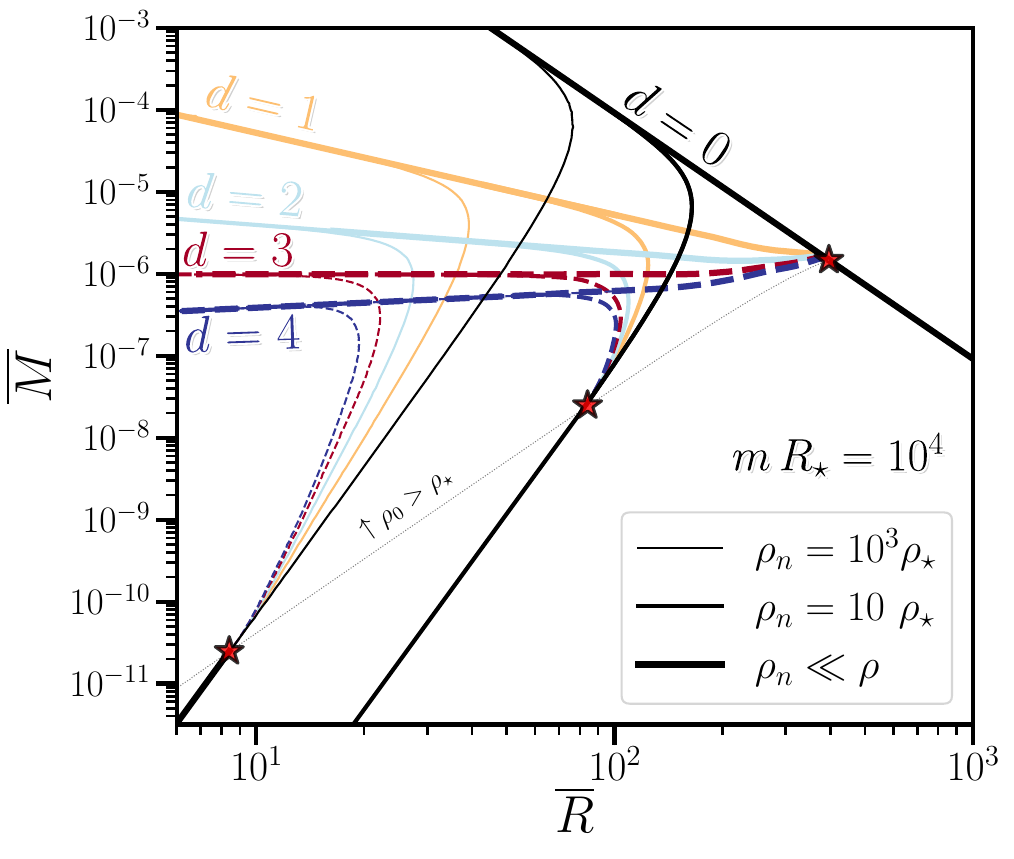}
             \caption{\justifying Mass--Radius relation obtained by solving 3D hydrostatic equations with XD adiabatic indices.  The $\textcolor{red}{\bigstar}$ marks values of central density, $\rho_0=\rho_{\star}$ (eq.~\eqref{eq:rhostar}). The thick curves correspond to configurations in which DM cloud self-gravitates (case I). The thin lines correspond to cases in which DM cloud is non-self gravitating, dominated by the background neutrons (case II). Here, $\bar{M}=M/\left(M_{\text{Pl}}^3/m^2\right)$ and $\bar{R} =R/\left(M_{\text{Pl}}/m^2\right)$.}
        \label{fig:M-R_LM_d}
\end{figure}

\noindent {\bf  I. Self-gravitating configurations ($\rho_n < \rho$) :} If the DM cloud is self-gravitating, the hydrostatic equations for a 3D spherical configuration take the form
\begin{align}\label{eq:hydro_sg}
  \frac{d\bar{\rho}}{d\bar{r}}&=-\frac{1}{c_s^2}\frac{1}{\bar{r}^2}\bar{\rho}\left(\bar{r}\right)\, \bar{M}\left(\bar{r}\right)~,\\
    \frac{d \bar{M}}{d\bar{r}}&=4\pi \bar{r}^2\bar{\rho}\left(\bar{r}\right)~.
\end{align}
We solve the above with the EoS dependent on the number of XD. Limiting analytical solutions are discussed in the appendix~\ref{app:LM_sols}. The resulting mass-radius configurations are shown in Fig.~\ref{fig:M-R_LM_d} (thick solid lines), for  $d=1$ (orange), $d=2$ (cyan), $d=3$ (red) and $d=4$ (blue) all for $\mdm R_\star = 10^4$. Starting from $\bar R \sim 10^3$, we can track the evolution of the configurations as the number of DM particles (and thus the mass $\bar M$) increases. At central densities above the critical density, $\rho_\star$, DM particles begin to occupy extra-dimensional phase space.  The corresponding critical mass, indicated by upper right $\textcolor{red}{\bigstar}$  in Fig.~\ref{fig:M-R_LM_d}, is given by
\begin{equation}\label{eq:Mcrit_sg}
M^{\rm SG}_{\text{crit}}\simeq 8 \frac{M_{\text{Pl}}^3}{m^2}
\frac{\rho_\star^{1/2}}{m^2}  \quad {\rm for} \quad \rho_{\star} \gg \rho_n~.
\end{equation}

Beyond $\rho_{\star}$, the EoS becomes very soft  and, for the same mass, the configurations for $d =1,2$ are much more compact  than for the $d=0$ case (black lines).  
For $d\geq 3 $, the critical mass turns out to be a maximum. Thus, if the captured mass is larger than this critical value, conditional to DM thermalization with NS, one expects that the DM sphere is unstable. We explicitly check this through a linear stability analysis in section~\ref{sec:instability}. The outcome, regardless of the higher dimensional gravitational pull, is the formation of a BH. Notice that the instability arises while the DM particles are still NR.\\

\noindent{\bf II. Non-self-gravitating configurations  ($\rho_n > \rho$) :} If the DM cloud is dominated by the gravitational potential of approximate constant density background neutrons, the hydrostatic equations reduce to, 
\begin{align}\label{eq:hydro_nsg}
  \frac{d\bar{\rho}}{d\bar{r}}&=-\frac{1}{c_s^2(\bar{r})}\frac{1}{\bar{r}^2}\bar{\rho}\left(\bar{r}\right)\times \frac{4}{3}\pi \bar{r}^3\bar{\rho}_n~,\\
    \frac{d \bar{M}}{d\bar{r}}&=4\pi \bar{r}^2\bar{\rho}\left(\bar{r}\right)~.
\end{align}
As in case I, the speed of sound is derived from the effective EoS.  The numerical solutions correspond to the thin solid lines in fig.~(\ref{fig:M-R_LM_d}). Asymptotically, the solutions are analytical  and scale as $\bar{M}(\bar{R}) \propto \bar{R}^6$.  At densities $\rho\ll m_{\text{DM}}/(3\pi^2R_\star^3)$, the equation of state is like that of $d=0$ degenerate fermions. At high densities, $i.e.$ $\rho\gg m_{\text{DM}}/(3\pi^2R_\star^3)$, the system behaves as a polytropic gas and the above equations can be solved analytically. For $d=3$, in terms of the central density $\bar{\rho}_0$, the solution is
\begin{eqnarray}\label{eq:mr_nonself}
    \bar{\rho}(\bar{r}) &=& \bar{\rho}_0 \left(1 - \frac{2\pi}{3^{4/3}}    \frac{m R_\star \bar{\rho}_n}{ \bar{\rho}^{1/3}_0}  \bar{r}^2\right)^{3}~, \\
    \bar{R} & =&      \frac{3^{2/3}}{\sqrt{2\pi}} \left(\frac{\bar{\rho}^{1/3}_0}{m R_\star \bar{\rho}_n} \right)^{1/2},\\
    \bar{M}(\bar{R}) &= &   \frac{16\sqrt{2\pi}}{35\pi} \left(\frac{\bar{\rho}_0}{m R_\star \bar{\rho}_n} \right)^{3/2} ~.\label{eq:m_nonself}
\end{eqnarray}
This solution of the mass approaches eq.~\eqref{eq:mass} as $\rho_0 \rightarrow \rho_n/2$. General solutions to arbitrary $d$ will be presented in a future work~\cite{GKTV:2026}.\\

As above, we can track the evolution of the configurations as the number of DM particles (and thus the mass $\bar M$) increases. When the density of the DM cloud reaches $\rho \sim \rho_{\star}$, DM particles propagate in XD phase space and the EoS becomes softer. The critical mass is now given by
\begin{align}\label{eq:Mcrit_nsg}
M^{\rm NSG}_{\text{crit}}=\frac{9\sqrt{3}\pi^{3/2}}{64}\frac{M_{\text{Pl}}^3}{m^2} \frac{\rho_\star^2}{\rho_{n}^{3/2}m^2}     \quad {\rm for} \quad \rho_\star \ll \rho_n~.    
\end{align}
As the central density further increases, the configurations become more compact, eventually reaching the solution of the self-gravitating branch when $\rho_\star\ll\rho_n\ll\rho$. This feature is observed for all $d\geq 0$. 
Unlike the self-gravitating scenario, there is no maximum stable mass for $d\geq 3$. Instead, the solutions are unstable to perturbations once the central density exceeds $\rho_{\star}$, see sec.~\ref{sec:instability}. Therefore, beyond such  points, marked by $\textcolor{red}{\bigstar}$ in Fig.~\ref{fig:M-R_LM_d}, there are no hydrostatic solutions. This, however, does not imply collapse to a BH. As there is no mechanism to evacuate energy, the sphere can not coherently shrink to the center and, instead there should be turbulence or radial mixing of layers resulting in an inhomogeneous DM sphere. As it continues to accumulate DM particles, it may eventually become self-gravitating and collapse into a BH. However, XD gravitational effects should probably be taken into account to draw definitive conclusions. \\

\subsection{Instability of the dark matter cloud}\label{sec:instability}
The stability of hydrostatic configurations can be ascertained by considering the pressure-averaged adiabatic index~\cite{book:shapiro,Chavanis:2006pf}
\begin{align}
    \left\langle\Gamma\right\rangle=\frac{\int d^3r P\left(r\right)\frac{d\log P}{d\log \rho}\left(r\right)}{\int d^3r P\left(r\right)}~.
\end{align}
The radial dependence of the pressure $P$ and the local polytropic index $\frac{d\log P}{d\log \rho}$ are extracted from the solutions of the hydrostatic equations, eqs.~\eqref{eq:hydro_sg} and~\eqref{eq:hydro_nsg}. Values of the average polytropic index $ \left\langle\Gamma\right\rangle \leq 4/3$ hint at instability against radial perturbation. We have performed linear stability analysis for all cases considered in this work, a detailed discussion will be provided in a future work~\cite{GKTV:2026}. 
\begin{figure}[htb!]
\centering
\includegraphics[width=0.6\linewidth]{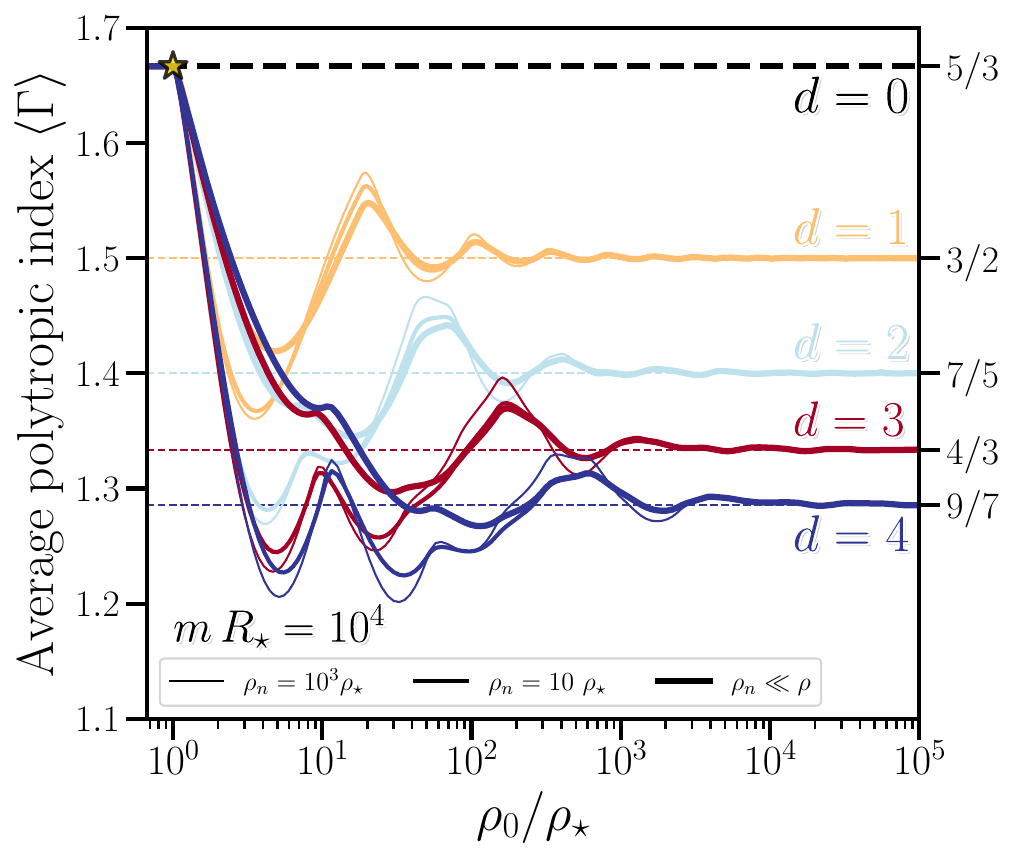}
    \caption{\justifying Averaged adiabatic index as a function of central density of hydrostatic configurations corresponding to fig.~(\ref{fig:M-R_LM_d}). The $\bigstar$ marks the critical density $\rho_{\star}$, eq.~\eqref{eq:rhostar}.}
    \label{fig:avgGamma}
\end{figure}

In Fig.~\ref{fig:avgGamma}, we report the pressure-averaged polytropic index for the $M-R$ configurations shown in Fig.~\ref{fig:M-R_LM_d}, as a function of the central density. For $d=3$, this criteria indicates that the collapse of self-gravitating configurations (thickest line) occurs at central densities of order $\rho_0\approx 20 \rho_{\star}$. However, from the existence of maximum mass, as seen in the mass-radius relationships, it is clear that collapse occurs at $\rho_0\approx \rho_\star$. This indicates that the  condition $\langle \Gamma \rangle \leq 4/3$ gives the correct answer within an order of magnitude. For non-self-gravitating configurations (thinner lines, $\rho_n \gg \rho_0$), instability occurs at central densities closer to the critical density ($\rho_\star$). In summary, the above results show that instability is expected to start once the threshold of the critical density is reached.\\

\noindent{\bf Collapse time scale:} For unstable configurations, perturbations grow with a time constant $\tau\sim R/c_{s}$ (see, ref.~\cite{book:shapiro}, chapter 6). Here, $R$ is the radius of the configuration and $c_{s}$ is the speed of sound. The instability timescale is evaluated at the critical density through  the speed of sound
\begin{align}
    c_{s,\text{crit}}=\sqrt{\frac{dP}{d\rho}}=\sqrt{\frac{\pi^{4/3}}{3^{1/3}}\left(\frac{\rho_\star}{m^4}\right)^{2/3}}~,
\end{align}
and through the expression for the radius of the star found from the hydrostatic equations
\begin{align}
R_{\text{crit}}=\frac{M_{\text{Pl}}}{m^2}\begin{cases}
 2.3\, \left(\frac{m^4}{\rho_{\star}}\right)^{1/6}    & \rho_{\star} \gg \rho_n\\
\frac{3^{5/6}\pi^{1/6}}{2}\left(\frac{\rho_n}{m^4}\right)^{1/2}\left(\frac{\rho_{\star}}{m^4}\right)^{1/3}  & \rho_{\star} \ll \rho_n~.\\     
\end{cases}
\end{align}
The instability timescale then reads
\begin{align}
\tau\simeq\begin{cases}
  2\frac{3^{1/6}}{\pi^{2/3}}\frac{M_{\text{Pl}}}{\sqrt{\rho_{\star}}} & \rho_{\star} \gg \rho_n\\
  \frac{3}{2\sqrt{\pi}}\frac{M_{\text{Pl}}\sqrt{\rho_n}}{m^4}& \rho_\star \ll \rho_n~.\\     
\end{cases}
\end{align}
Note that, in the self-gravitating case, the time scale is independent of $\rho_0$ and depends on $R_\star$ through $\rho_{\star}$ and collapse is ensured when $\rho_{\star} \gtrsim \rho_n$, along with $\langle \Gamma \rangle \leq 4/3$. For the non-self-gravitating case, however, the interpretation of the hydrostatic instability is not fully clear. Comparing the instability time scale with the free fall time, $t_{\rm free} \sim (G \rho_n)^{-1/2}$, we find that $t_{\rm free} < \tau$, for $m \lesssim 0.2\,{\rm GeV}$. Such an inequality might indicate collapse to a BH, as for the self-gravitating case. However, we suspect that the unstable mode actually signals the emergence of  convective instability, leading to radial turbulent mixing. It is thus likely that for larger masses ($t_{\rm free}> \tau$), the configuration quasi-statically adjusts itself without collapsing to the center. This  can in principle proceed until the DM cloud becomes self-gravitating. For $d \geq 2$, the DM cloud should collapse into a BH. 
As this neglects the modification of the gravitational interaction, which should affect the core of the DM cloud, these results for the non-self-gravitating case are tentative and clearly deserve further investigation \cite{GKTV:2026}.

\section{Dark Matter Capture and Thermalization in Neutron Stars}\label{sec:cap_therm}
\label{sec:DMconstraints}

So far, we have described the hydrostatic configuration of the dark matter sphere, assuming it is initially in thermal equilibrium with the neutron background. We now turn to the mechanism by which dark matter particles are captured by the neutron star from the surrounding halo.

We consider a typical neutron star of mass $1.5 \, M_\odot$ and radius $R_{\rm NS} = 12 \,{\rm km}$. The geometric capture rate of DM particles, for local DM density of $\rho = 0.4\,{\rm GeV/cm^3}$, velocity dispersion $v_d = 270\, {\rm km/s}$ is~\cite{Garani:2018kkd}
\begin{eqnarray}\label{eq:capg}
  C^{\rm geom} &=& 4\times10^{24} {\rm s^{-1}} \left(\frac{\rho}{0.4\,{\rm GeV/cm^3}} \frac{\rm GeV}{\mdm}\right)~.
\end{eqnarray}
The maximum capturable mass over the NS life span ($t_{\rm NS} = 10\, {\rm Gyr}$ ) is the geometric mass $M^{\rm geom}_{\rm cap} = \mdm\, C^{\rm geom}\,t_{\rm NS} = 10^{42}\,{\rm GeV}$.
The capture rate in the thin regime is~\cite{Gould:1987ir,McDermott:2011jp}
\begin{eqnarray}\label{eq:capw}
  C &\approx& 10^{22} \,{\rm s^{-1}} F(\mdm) \left(\frac{\rm TeV}{\mdm}\right)\left(1- \frac{1-e^{-A^2}}{A^2} \right) \left(\frac{\sigma_{\chi n}}{10^{-45}\,{\rm cm^2}} \right)~.
  \end{eqnarray}
where $A^2 = 6 \mdm m_n v^2_{\rm esc}/(v_d^2 (\mdm -m_n))$, and $F(\mdm) = {\rm Min(1, \sqrt{2} \mu_r v^2_{\rm esc}/E^n_F)}$. The saturation cross section is set by  $C=C^{\rm geom}$. The corresponding captured mass over course of life time of NS is $M_{\rm cap} = \mdm\, C\, t_{\rm NS}$.

After DM capture, DM particles are mostly in radial orbits, losing energy in each collision with background neutrons. Eventually DM particles reach thermal equilibrium with the NS matter. The time taken to thermalize with NS medium is given by~\cite{Bertoni:2013bsa, Garani:2020wge} 
\begin{equation}\label{eq:ttherm}
t_{\rm therm} = 10700\,{\rm yrs}\,\frac{\mdm m_n}{(\mdm+m_n)^2}\left(\frac{10^5\,{\rm K}}{T_{\rm NS}} \right)^2 \frac{10^{-45}\,{\rm cm^2}}{\sigma_{\chi n}}~.
\end{equation}
For very large DM masses ($\gtrsim 10^8\,{\rm GeV}$), we use the thermalization time given by a qualitatively different expression~\cite{Bell:2023ysh}, 
\begin{equation}
    t_{\rm therm} \,\sim\,{\rm Gyrs} \,\frac{\mdm}{10^9\,\rm GeV} \frac{2\times10^{-57}\, \rm cm^2}{\sigma_{\chi n}}
    \log\frac{m}{T_{\text{NS}}}~.
\end{equation}

\section{Black hole formation, accretion and evaporation}\label{sec:BH_stuff}  
 We have illustrated in sections~\ref{sec:hydro} and \ref{sec:instability}, that there exists a critical DM mass beyond which hydrostatic instability sets in, leading to BH formation. In this section, we will discuss the evolution of the formed BH. This includes accretion of neutron matter by BH and their evaporation.
 
We have shown that after $M^{\rm SG}_{\text{crit}}$ (see eq.~\eqref{eq:Mcrit_sg}) amounts of DM has been accreted (and thermalized), the cloud is destabilized and collapses. First, to ensure that a horizon forms, the collapsing mass $M^{\rm SG}_{\text{crit}}$ should be at least larger than the XD Planck mass~\cite{Myers:1986un, Argyres:1998qn}, 
\begin{align}\label{eq:horizon}
M^{\rm SG}_{\text{crit}}>M_{\text{Pl}}^\star=\left(\frac{M_{\text{Pl}}^2}{R_\star^d}\right)^{\frac{1}{2+d}}~.
\end{align}
If this is satisfied,  a BH of mass $M_{\text{BH}}=M^{\rm SG}_{\text{crit}}$ is formed, with horizon radius~\cite{Myers:1986un, Argyres:1998qn}
\begin{align}
  R_{\rm BH} = \left(\frac{8\Gamma(\frac{d+3}{2})}{d+2} \frac{(2\pi R_\star)^d}{(\pi^{(d+1)/2})}\frac{M_{\rm BH}}{M^2_{\rm Pl}} \right)^{\frac{1}{1+d}}~.
\end{align}
If $R_{\rm BH} \lesssim  R_\star$, the BH is higher dimensional, while it is effectively 3D  if  $R_{\rm BH} > R_\star$. The corresponding critical mass  is 
\begin{align}
     M_{\text{BH}}^{\text{fit}}&=\frac{(2+d)\sqrt{\pi}}{2^{d+3} \Gamma\left(\frac{3+d}{2}\right)}M_{\text{Pl}}^2 R_\star \overset{d\rightarrow 3}{=} 5\times 10^{37}\left(\frac{R_\star}{\text{fm}}\right)\text{ GeV}.
\end{align}
Equivalently, a black hole lighter than $M_{\text{BH}}^{\text{fit}}$ is effectively higher-dimensional. Otherwise, it is effectively a 3D BH. This distinction is important to estimate the accretion rate of neutron matter onto the BH and the amount of mass it loses through Hawking evaporation, 
\begin{equation}\label{eq:whowins}
  \frac{d M_{\rm BH}}{d t} = \frac{dM_{\text{BH}}^{\text{accr}}}{dt}
+\frac{dM_{\text{BH}}^{\text{evap}}}{dt}~.
\end{equation}

\subsection{Accretion} 
Accretion of matter by the BH is proportional to the cross section. The accretion rate can be generally written as
\begin{align}\label{eq:accre}
    \frac{dM_{\text{BH}}}{dt}^{\text{accr}}=\sigma_{\text{accr}}\, \rho_n \, v ,
\end{align}
where $\sigma_{\text{accr}}$ is an accretion cross-section and $\rho_n$, $v$ are the energy density and velocity of the neutron matter in the background. We assume $v\approx 0.5$ and $\rho_n\approx 10^{15}\,{\rm g/cm^3}$.\\

Independent of the dimensionality of the BH, there are two distinct regimes of accretion~\cite{Giffin:2021kgb}: hydrodynamical (Bondi) when $\lambda_{\rm dB}=1/(m_n v) \lesssim R_{\rm BH}$ and $R_{\rm BH} > \lambda_{\rm mfp} c^2_s$. Quantum (Unruh)  when $\lambda_{\rm dB} \gtrsim R_{\rm BH}$ and $ R_{\rm BH} <  \lambda_{\rm mfp} c^2_s$. In the Bondi regime, we have~\cite{Bondi:1952}
\begin{align}
    \sigma_{\text{accr}}^{\text{Bondi}}=\frac{\pi R_{\text{BH}}^2}{v^4}.
\end{align}
While in the Unruh regime diffractive effects are dominant. Therefore, the correct resummed partial wave cross section should be considered in the estimate~\cite{Unruh:1976},
\begin{align}
\sigma_{\text{accr}}^{\text{Unruh}}&=\frac{2\pi R_{\text{BH}}^2}{v}\frac{\xi}{1-e^{-\xi}}\\
    \xi&=\pi\frac{1+v^2}{v^2\sqrt{1-v^2}}\frac{R_{\text{BH}}}{\lambda_{\text{DB}}}.
\end{align}

For $d =0$, quantum accretion is relevant important when $M_{\rm BH} \lesssim M^2_{\rm pl}/m_n  = 10^{38}\,{\rm GeV}$. Similarly for $d=3$, when $M_{\rm BH} \lesssim M^2_{\rm pl}/(m_n^4 R_\star^3) =2\times10^{35} \,{\rm GeV} ({\rm fm}/R_\star)^3$. In the intermediate regime, $\lambda_{\rm dB} <  R_{\rm BH}$ and $ R_{\rm BH} < \lambda_{\rm mfp} c^2_s$, the classical absorption cross sections should be used~\cite{Giffin:2021kgb}.\\

\subsection{Evaporation}  The rate of evaporation for a 3+d dimensional BH is~\cite{Johnson:2020tiw} 
\begin{equation}\label{eq:bh_evap}
    \frac{d M_{\rm BH}}{d t} = -P_{\rm brane}(M_{\rm BH}) - P_{\rm bulk}(M_{\rm BH})~.
\end{equation}
The blackbody temperature is related to the horizon size, given by $T_{\rm BH} = (1+d)/(4\pi R_{\rm BH})$. 
Emission in the bulk is $P_{\rm bulk} \sim g_{eff} R^{2+d}_{\rm BH} T^{4+d}_{\rm BH}\sim R^{-2}_{\rm BH}$. Emission in the brane is similar upto geometric and gray body factors, $P_{\rm brane} \sim c_{eff} R^{-2}_{\rm BH}$.  The effective degrees of freedom $g_{eff} (c_{eff})$ can change if the emitted particle is a boson or a fermion~\cite{Page:1976df,Page:1976ki,MacGibbon:1991}. For bosons (fermions) the constant is $\propto \zeta(d+4) \,(\eta(d+4))$, Riemann zeta (Dirchlet eta)~\cite{Johnson:2020tiw}. Therefore, the evaporation rate can be written as
\begin{align}
&\frac{dM_{\text{BH}}^{\text{evap}}}{dt}\approx- \mathcal{C}_d \left(\frac{M_{\text{Pl}}^2}{R_\star^d}\right)^{\frac{2}{1+d}} \left(\frac{\text{GeV}}{M_{\text{BH}}}\right)^{\frac{2}{1+d}}\text{GeV}^2\label{eq:bh_evap_power} ~.
\end{align}
The coefficient $\mathcal{C}_d$ includes contribution from the KK modes but not standard model particles, 
\begin{align}\label{eq:bh_evap_const}
\mathcal{C}_d=\frac{\left(1+d\right)^{3}\left(2+d\right)^{\frac{2}{1+d}}\left(3+d\right)^{\frac{3+d}{1+d}}\pi^{\frac{2\left(1-d\right)}{1+d}}}{30720\, \Gamma\left(\frac{3+d}{2}\right)^{\frac{2}{1+d}}}~.
\end{align}

Finally, note that if the BH horizon is larger than the extra-dimension, $M_{\text{BH}}>M_{\text{BH}}^{\text{fit}}$, the evaporation rate is that of the case $d=0$. Considering the so-called memory burden effect in this scenario will increase the evaporation time~\cite{Dvali:2020wft,Alexandre:2024nuo,Dvali:2024hsb, Basumatary:2024uwo, Dvali:2025ktz}. We do not consider this for the current discussion. Thus, the exclusion limits obtained in this work are conservative.

If the evaporation term dominates (depending on the initial BH mass), the lifetime reads~\cite{Anchordoqui:2024dxu}
\begin{equation}
    t_{\rm evap} =  - \int^{M^{\rm fit}_{\rm BH}}_{M_{\rm BH}} \frac{d M_{\rm BH}}{P^{d=0}_{brane}} - \int^0_{M^{\rm fit}_{\rm BH}} \frac{d M_{\rm BH}}{P^{d=3}_{brane} + P^{d=3}_{bulk} }~. 
\end{equation}
The evaporation rates $P_{brane}$ and $P_{bulk}$ are given above in  eqs.~\eqref{eq:bh_evap}, \eqref{eq:bh_evap_power} and \eqref{eq:bh_evap_const}. \\

\section{\bf DM constraints from neutron star collapse} \label{sec:constraints}
\begin{figure}[htb]
\centering
\includegraphics[width=0.7\linewidth]{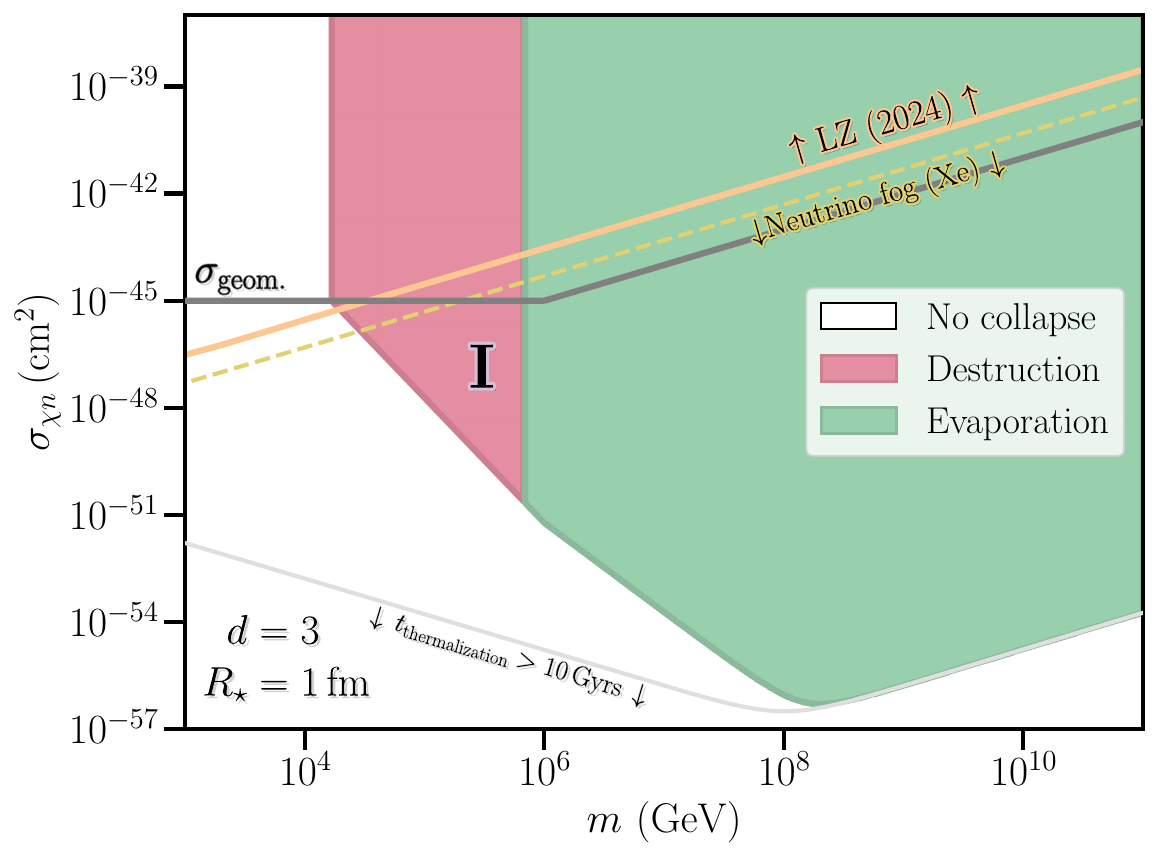}\\
\includegraphics[width=0.7\linewidth]{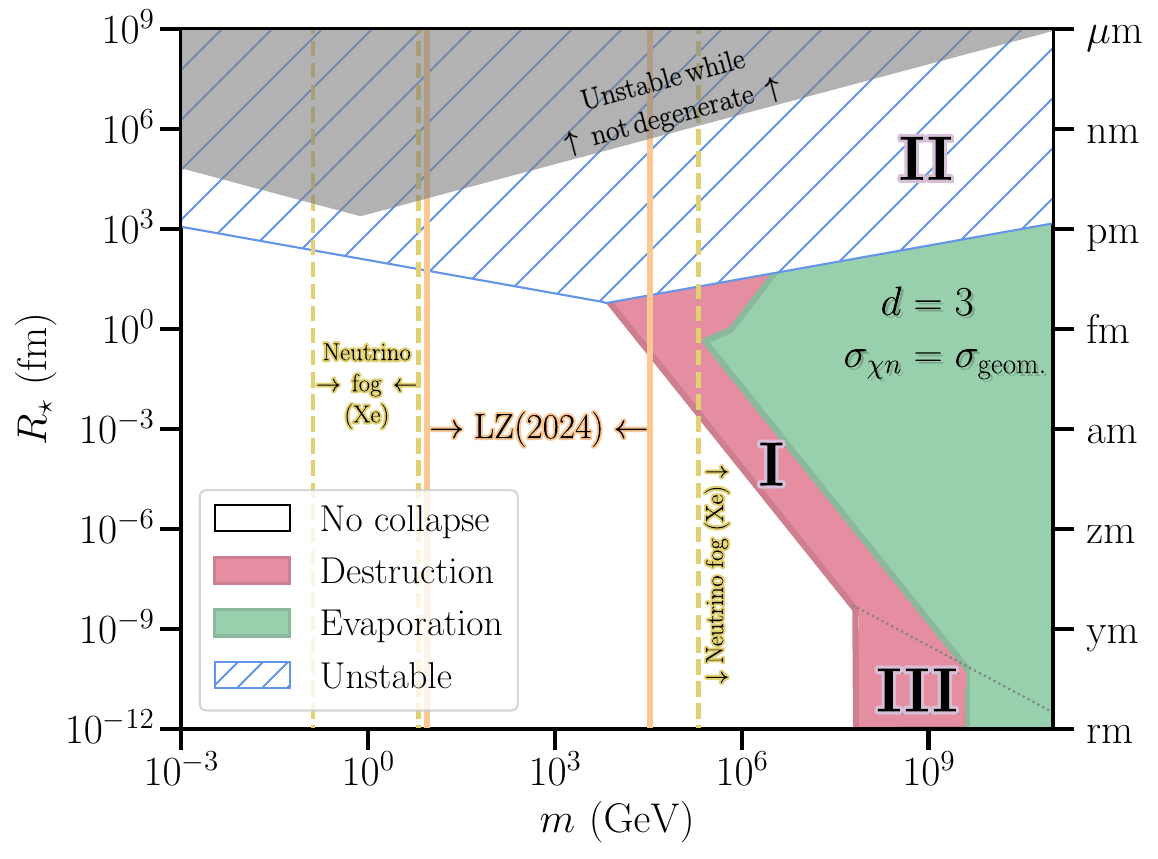}\\
\caption{\justifying Constraints from old NS for scenarios with three XD. In the red regions, the dark matter cloud is destabilized by propagation in XD, forming a BH that consumes the NS. Green regions correspond to BHs light enough to evaporate within $\rm Gyrs$. Solid yellow curves show current LZ limits~\cite{LZ:2024zvo}, while dashed yellow lines indicate the neutrino fog~\cite{OHare:2021utq}. White regions are unconstrained, either because collapse criterion is not met or DM does not thermalize with neutron matter. Region ${\bf I}$ corresponds to self-gravitating NR collapse; Region ${\bf II}$ 
corresponds to perturbatively unstable configurations but does not necessarily lead to collapse;
 and region ${\bf III}$ to relativistic collapse. \underline{\it Top panel:} DM–neutron cross section vs DM mass for $d=3$ and $R_\star =1~{\rm fm}$. \underline{\it Bottom panel:} For saturation cross section and $d=3$, constraints on $R_\star$ vs. DM mass; black shading marks parameters for which the DM cloud is unstable before onset of degeneracy.
}
\label{fig:CollapseParamSpace}
\end{figure}
Putting together all the ingredients from sections~\ref{sec:hydro},~\ref{sec:cap_therm}, and \ref{sec:BH_stuff}. We can now derive constraints on $\mdm$, $R_\star$ and DM-neutron scattering cross section $\sigma_{\chi n}$ assuming $d=3$, see fig.~(\ref{fig:CollapseParamSpace}). We outline the procedure below. First, for given $\sigma_{\chi n}$, the thermalization time, see eq.~\eqref{eq:ttherm}, should be less than the age of the NS, $t_{\rm therm} < t_{\rm NS}$. Second, enough DM particles should be captured from the halo to reach ${\rm Min(M^{\rm SG}_{\rm crit}, M_{\rm Ch}) }~<~M_{\rm cap}$, see section~\ref{sec:cap_therm}. If these conditions are met, the entire neutron star is transformed into a solar-mass scale BH (red region). Finally, we require that the BH formed does not evaporate fast, i.e. $t_{\rm evap} > t_{\rm NS}$. The regions of the parameter space where this is not satisfied are shaded \textcolor{teal}{green}, while in \textcolor{gray}{gray} shaded regions, instability occurs before DM becomes degenerate.

In the upper panel of Fig.~\ref{fig:CollapseParamSpace}, the excluded regions in $\mdm - \sigma_{\chi n}$ plane are shown for fixed  $R_\star=1~{\rm fm}$.  In the lower panel, the excluded regions in the $\mdm - R_\star$ plane are shown for geometric accretion rate, corresponding to saturation cross section.  Also shown in solid yellow curves are the current LZ limit on spin-independent scattering cross section~\cite{LZ:2024zvo}, and dashed yellow curves indicate the neutrino floor~\cite{OHare:2021utq}.\\

The DM cloud collapses when NR, if $M^{\rm SG}_{\rm crit} < M_{\rm cap}$. The mass dependence results from the capture rate, eq.~\eqref{eq:capw}, and the parametric form of eq.~\eqref{eq:Mcrit_sg}.  We distinguish: 

\noindent{(I)} Self-gravitating DM cloud:  When the condition $M^{\rm SG}_{\rm crit} < M_{\rm cap}$ is satisfied, collapse occurs when the DM cloud is self-gravitating and the DM NR.  For a NS, the lightest DM that can self-gravitate is few TeV.
The limits on $\bar \sigma_{\chi n} = \sigma_{\chi n}/10^{-45}$ cm$^2$ are:
\begin{align}
  {\bar \sigma_{\rm \chi n}}& \gtrsim
  \begin{cases}
    \left(\frac{\rm TeV}{m} \right)^{7/2} \left(\frac{\rm 700\, fm}{R_\star} \right)^{3/2} & { \rm TeV} < m < {\rm PeV} \\
    \left(\frac{\rm PeV}{m} \right)^{5/2} \left(\frac{ 7 \times 10^{-5}
    \rm  fm}{R_\star} \right)^{3/2} &  m > {\rm  PeV} 
    \end{cases}
  \end{align}

\noindent{(II)} Non-self-gravitating DM cloud:  Hatched regions mark parameter space where the condition $M^{\rm NSG}_{\rm crit} < M_{\rm cap}$ is satisfied. While hydrostatic configurations are unstable, collapse to a BH does not necessarily follow. However, as more DM particles are accreted, the density increases and the critical mass for collapse to BH approaches eq.~\eqref{eq:Mcrit_sg} as $\rho \gtrsim \rho_n$. A  discussion of this regime, taking into account extra-dimensional gravity, is the object of work in progress~\cite{GKTV:2026}.\\

\noindent{(III)} Relativistic DM: The cloud collapses when DM particles become relativistic, $\rho \gg \mdm^4$. The condition for collapse is then $M_{\rm Ch}  < M_{\rm cap}$,  and $\mdm \gtrsim 0.06\,{\rm GeV}\,({\rm fm}/R_\star)$.\\

\section{Conclusions}\label{sec:conclusion}
The existence of old neutron stars imposes strong constraints on DM–nucleon cross sections and mass, and on the size and number of extra dimensions (XD). We have shown that, for geometric cross sections, fermionic asymmetric DM with masses above $\mathcal{O}(10\,{\rm TeV})$ is excluded when $d\geq 3$
extra dimensions are accessible. Neutron stars thus serve as sensitive probes of both dark sectors and spacetime dimensionality, motivating further exploration of compact-object signatures in extra-dimensional setups.

Several simplifying assumptions underlying these results deserve to be stated explicitly. In particular, we have treated the problem as being purely 3 dimensional from the perspective of gravity, only taking into account the modification of the equation of state as DM particles start to fill the extra-dimensional momentum space. This is a dramatic simplification, which circumvents the complication posed by the lack of symmetry when going from a DM cloud in $D=3$ to one that has a core that can start filling the $D=3+d$ spatial dimensions. At the same, this very simplification allows to make robust statements regarding the fate of the DM cloud, at least for $d \geq 2$ extra dimensions, as depicted in Figs. \ref{fig:M-R_LM_d} and \ref{fig:CollapseParamSpace}.
Our next goal is to go beyond this approximation and try to take into account the modification of gravity \cite{GKTV:2026}. As argued throughout this work, this should facilitate collapse, thus possibly enlarging the constraints on DM parameters. This would also allow us to treat more definitively the non-self-gravitating regime part of our analysis, which potentially concerns a broad range of DM particle masses (see hatched region II in Fig.\ref{fig:CollapseParamSpace}, lower panel). For this regime, the impact of extra-dimensional gravitational interactions on the hydrostatic structure of the DM cloud must be fully accounted for.  

On the astrophysical side, we have adopted standard assumptions for neutron star properties and DM capture rates; uncertainties in the local DM density, DM velocity distribution, and neutron star internal structure could affect the precise location of the exclusion boundaries, though we expect the qualitative picture to be robust.
Finally, other celestial objects such as white dwarfs or the Sun may provide complementary constraints in different regions of DM parameter space~\cite{GKTV:2026}, and a systematic comparison across object types would sharpen the overall picture.

On the particle physics side, we notice that, in the regime of self-gravitating DM (regime I in Fig. \ref{fig:CollapseParamSpace}, lower panel), heavy DM particles within neutron stars can probe very small extra dimensions. Throughout, we have made the starting assumption that only the DM particles can probe the extra dimensions. For $R^{-1} \gtrsim 1$ TeV, or $R \lesssim 10^{-4}$ fm, which is the realm of Universal Extra Dimensions scenarios (UED) \cite{Appelquist:2000nn}, this assumption can be clearly relaxed, allowing all particles of the Standard Model, or rather their KK modes, to explore the extra dimensions. By KK parity, the UED framework naturally predicts DM candidates, albeit in the form of the KK partner of the hypercharge gauge boson \cite{Servant:2002aq}. Whether a self-consistent DM scenario, including asymmetric DM fermions, can be developed in some kind of UED framework is a question we leave for possible future work.

\section*{Acknowledgments}
The work of RG is supported by ANRF/ECRG/2025/000418/PMS. 
The work of JV was supported by the Collaborative Research Center SFB1258 and by the Deutsche Forschungsgemeinschaft (DFG, German Research Foundation) under Germany’s Excellence Strategy - EXC-2094 - 390783311 and by Universität Bielefeld.
The work of MHGT is supported by the belgian FRS/FNRS and the IISN convention No. 4.4503.15.

\bibliographystyle{JHEP-mod}
\bibliography{biblio}
\appendix
\section{Equation of state for degenerate Fermi gas} \label{app:xdims0}

Fermionic DM particles are being continually added to neutron stars living in 3-spatial dimensions. These fermions begin to occupy states in extra dimensions as soon as their Fermi energy is of the order of first mode excitation energy. The 3D number density, energy density and pressure for such particles read as follows, 
\begin{eqnarray}\label{eq:thermo_d}
    n &=&  g \sum_{p_\perp} \int \frac{d^3 p}{(2 \pi)^3} \frac{1}{e^{\beta(E-\mu)} +1}~, \\
     \rho &=&  g \sum_{p_\perp} \int \frac{d^3 p}{(2 \pi)^3} \frac{E}{e^{\beta(E-\mu)} +1}~, \label{eq:thermo_rho}\\
     P &=&  \frac{g}{3} \sum_{p_\perp} \int \frac{d^3 p}{(2 \pi)^3} \frac{p^2}{ E}\frac{1}{e^{\beta(E-\mu)} +1}~.\label{eq:thermo_p}
\end{eqnarray}
In the non-relativistic limit, $E -\mu \equiv p^2/2m + p^2_\perp/2m -\bar{\mu}$. The momenta perpendicular to SM brane is $p_\perp^2 =\sum_d  n^2_d/R^2_d$. The number of compactified dimensions is denoted by $d$, and their size by $R_d= R_\star/(2\pi)$. Retaining the temperature dependence, we perform the 3-momenta integrals, yielding the following expressions for energy density and the pressure:

\begin{eqnarray}\label{eq:rhop_d}
\frac{\rho}{m} \left(\frac{2\pi}{m T} \right)^{3/2} &=& - \sum^{+\infty}_{n_d=-\infty} \left( {\rm Li_{3/2}(-z_{n_d})} + \frac{3 T}{2 m}{\rm Li_{5/2}(-z_{n_d})}\right)~, \\
   \frac{P}{T} \left(\frac{2\pi}{m T} \right)^{3/2} &=& - \sum^{+\infty}_{n_d=-\infty}\left( {\rm Li_{5/2}(-z_{n_d})} + \frac{5 T}{2 m}{\rm Li_{7/2}(-z_{n_d})} \right)~, \label{eq:Pp_d} \\
 z_{n_d} &=& \exp\left(-\sum _d \frac{n^2_d}{2 \, m  T R_d^2} + \frac{\bar{\mu}}{T}\right)~.\label{eq:Z_d}
\end{eqnarray}
Information about the extension of Fermi-sphere is contained in the parameter $z_{n_d}$. Note that the sum over number of extra-dimensions is implicit and suppressed in the above expression. Dimensionless energy density and pressure are defined as $\rho^\prime  \equiv \frac{\rho}{m} \left(\frac{2\pi}{m T} \right)^{3/2}$ and $P^\prime \equiv \frac{P}{T} \left(\frac{2\pi}{m T} \right)^{3/2}$.

\begin{center}
    \begin{figure*}[htb!]
        \centering
\includegraphics[width=0.49\linewidth]{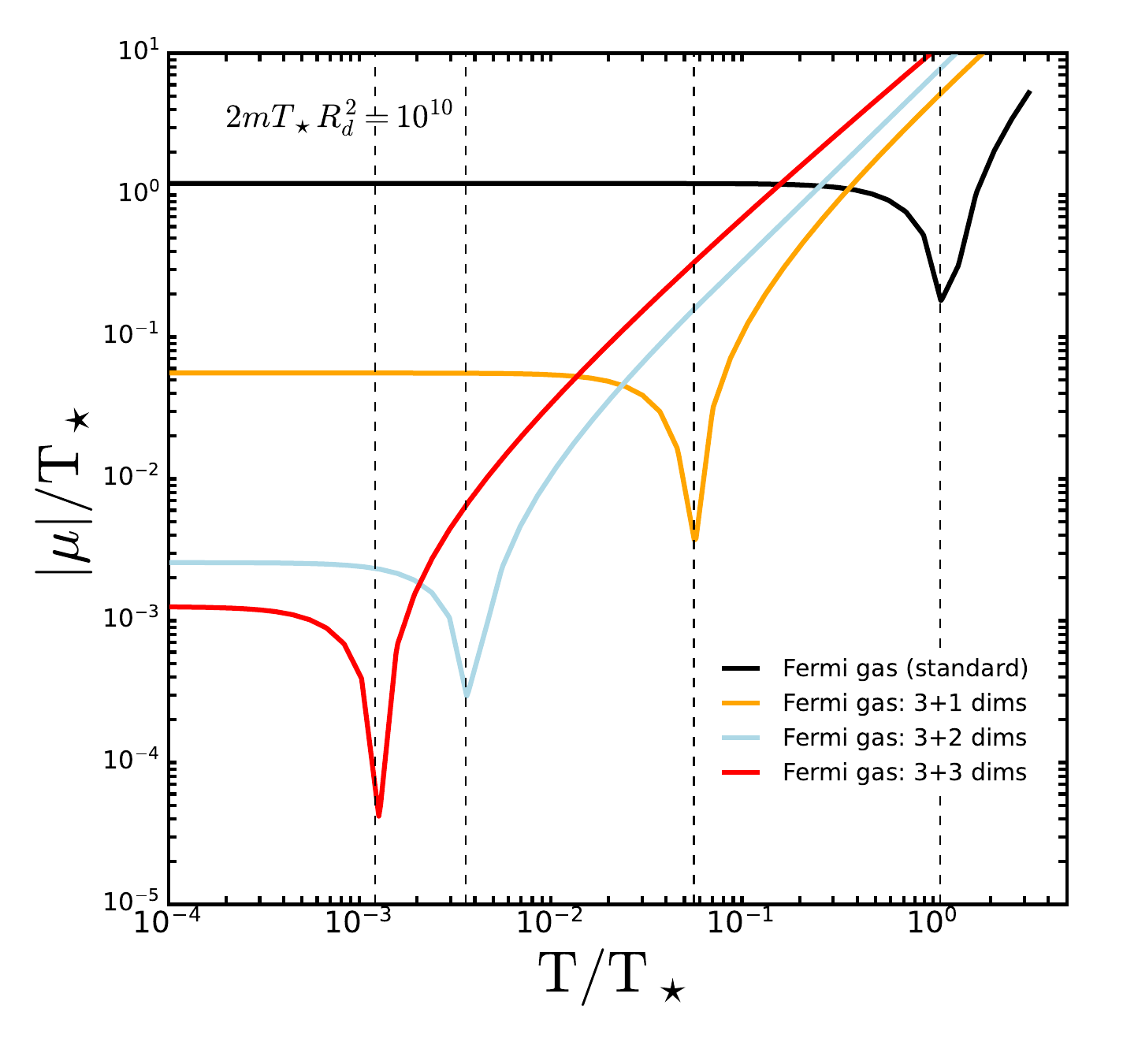}                \includegraphics[width=0.49\linewidth]{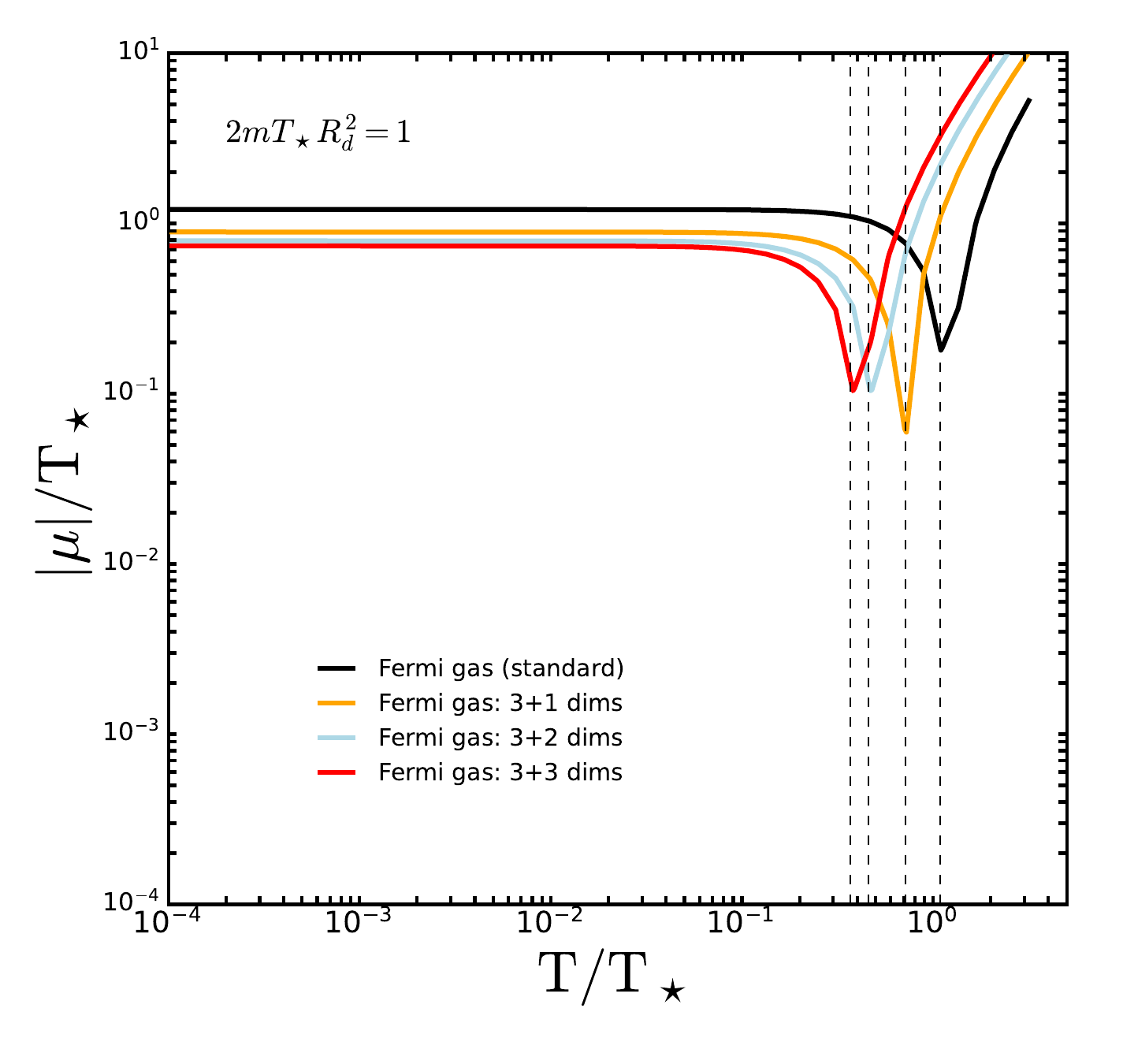}
        \caption{Chemical potential versus temperature at constant number density ($N/V_3$). }
        \label{fig:muT}
    \end{figure*}
\end{center}
A qualitative understanding can be reached by examining the behavior of the  chemical potential as a function of temperature at fixed number density~\cite{FetterWalecka:1971,Kapusta:2006pm}. In the non-relativistic limit the number density is simply $\rho/m$. This forumla is readily available, given by eq.~\eqref{eq:rhop_d}. 

In $D=3$ spatial dimensions, the familiar Maxwell–Boltzmann expression
for number density is 
\begin{equation}
    n=N/V_3 = g (m T/2\pi)^{3/2} e^{\mu/T}~.
\end{equation}
This may be inverted to the  form 
\begin{equation}
    \mu = T \log(n \lambda^3)~,
\end{equation}
where the thermal wavelength $\lambda = \sqrt{2\pi/(m T)}$ dictates the scale at which the chemical potential crosses from negative to positive. The criterion $n \lambda_\star = 1$
 thus marks the on-set of degeneracy.

In Fig.~(\ref{fig:muT}) the behavior of $|\mu|/T_\star$ is shown as a function of $T/T_\star$ for a fixed 3D number density ($N/V_3$), for $2 m T_\star R_d^2 = 10^{10}$ (left panel) and $2 m T_\star R_d^2 = 1$ (right panel). In each of these plots we show the relation for standard 3D Fermi-gas (\textcolor{black}{black}), Fermi-gas in 3+1  (\textcolor{orange}{orange}), 3+2  (\textcolor{blue}{blue}), and 3+3 dimensions (\textcolor{red}{red}). Fermions become degenerate when the chemical potential becomes positive. This is seen to be a dip in Fig.~(\ref{fig:muT}) as we plot $|\mu|/T_\star$ in log-scale).

The behavior of these curves can be qualitatively understood by examining various limits of the eq.~\eqref{eq:rhop_d}. We first consider the large temperature limit,
\begin{eqnarray}
  n\left(\frac{2\pi}{m T_\star}\right)^{3/2}&& \stackrel{T\rightarrow \infty}{\sim} \sum_d \sum_{n_d}\left(\frac{T}{T_\star}\right)^{3/2}    \Bigg[\zeta(3/2) - 2\sqrt{\frac{\pi T_\star}{T}}\left(\frac{n^2_d }{2 m T_\star R_d^2} -\frac{\bar{\mu}}{T_\star}\right)^{1/2} \Bigg. \nonumber \\
  && \Bigg.+ \frac{T^2_\star}{T^2} \left(\frac{n^2_d }{2 m T_\star R_d^2} -\frac{\bar{\mu}}{T_\star}\right) \zeta(1/2) \Bigg]~.
\end{eqnarray}
At large temperatures ($T/T_\star$), $\mu$ tends to be negative. With extra dimensions they tend to be more negative. In other words, modes propagate in extra dimemsions provided $\lambda_\star/R_d \lesssim \mathcal{O}(1)$, as expected from the above equation. To summarize, while extra dimensions are relevant as soon as $R_d \gtrsim 1/(m T)^{1/2}$, at finite density they become  relevant for non-relativistic particles when $\lambda \lesssim 1/n^{1/3}$ (and thus quantum statistics). The greater the number of extra-dimensions, the lower will be the critical temperature. This is illustrated clearly in Fig.~(\ref{fig:muT}). The left panel correspond to instances when the size of extra-dimensions is larger than the critical de Broglie wavelength, i.e. extra-dimensional phase space is fully accessible. While in the right panel, the the critical de Broglie wavelength is of the order of the size of extra-dimensions. 

Finally we consider the degenerate limit, the particle distribution functions in eqs.~\eqref{eq:thermo_d},~\eqref{eq:thermo_rho} and ~\eqref{eq:thermo_p} reduce to Heaviside theta functions. The number density and the pressure have the form,
\begin{align}\label{eq:numden3d}
n &=\frac{g}{2\pi^2}\sum\limits_{d}\sum\limits_{\vec{n}_d=-\infty}^{+\infty}\int\limits_{0}^\infty dp p^2\theta\left(\bar{\mu}-\left(\frac{p^2}{2m}+\frac{\left|\vec{n}_d\right|^2}{2 m R_d^2}\right)\right)~.
\end{align}
The energy density is $\rho = m n$, and the effective 3D pressure is given by 
\begin{align}\label{eq:p3d}
P &=\frac{g}{2\pi^2}\sum\limits_{d}\sum\limits_{\vec{n}_d=-\infty}^{+\infty}\int\limits_{0}^\infty dp p^2 \frac{p^2}{3 E}\theta\left(\bar{\mu}-\left(\frac{p^2}{2m}+\frac{\left|\vec{n}_d\right|^2}{2 m R_d^2}\right)\right)~.
\end{align}
For fixed 3D number density $n$, the chemical potential $\bar{\mu}$ becomes smaller than $n^{1/3} $ due to the discrete sum over the number of states. This is precisely the behavior observed in Fig.~\ref{fig:muT}.

\subsection{Parametric behavior of the polytropic index} 

Explicitly performing the integrals in eq.~\eqref{eq:numden3d} and eq.~\eqref{eq:p3d}, we have

\begin{align}
n&=\frac{k_F^3}{3\pi^2}\left(1 + \sum\limits_{1\leq \left|\vec{n}\right|\leq \lfloor Y \rfloor}\left(1-\frac{\left|\vec{n}\right|^2}{Y^2}\right)^{3/2}\right),\\
P&=\frac{k_F^5}{15\pi^2m}\left(1 + \sum\limits_{1\leq \left|\vec{n}\right|\leq \lfloor Y \rfloor}\left(1-\frac{\left|\vec{n}\right|^2}{Y^2}\right)^{5/2}\right).
\end{align}
where we defined the Fermi momentum via $\bar{\mu}=\frac{k_F^2}{2m}$ and the dimensionless parameter $Y=R_d k_F$ controls the importance of the extra-dimension. In these expressions, the first term corresponds to the standard 3-dimensional results. The second term sums all allowed KK modes, that is up to the integer part of $Y$, written above in terms of the floor function $\lfloor Y \rfloor$. We will denote the sum over KK modes as
\begin{align}
    \mathcal{S}_\kappa\left(Y\right)\equiv \sum\limits_{1\leq \left|\vec{n}\right|\leq \lfloor Y \rfloor}\left(1-\frac{\left|\vec{n}\right|^2}{Y^2}\right)^{\kappa/2}\label{eq:Sk}
\end{align}
and the polytropic index can thus be directly computed
\begin{align}
    \gamma=\frac{d\log P}{d\log \rho}=\frac{5+\frac{Y}{1+\mathcal{S}_5\left(Y\right)}\frac{d\mathcal{S}_5}{dY}}{3+\frac{Y}{1+\mathcal{S}_3\left(Y\right)}\frac{d\mathcal{S}_3}{dY}}
\end{align}

When $Y<1$, that is $k_F<1/R_d$, the highest momentum modes do not have enough energy to explore the extra-dimension, the KK sum does not contribute and we recover the polytropic exponent of the standard free, degenerate Fermi gas, $i.e.$ $\gamma=5/3$.\\

\begin{figure}[h!]
    \centering
    \includegraphics[width=0.47\linewidth]{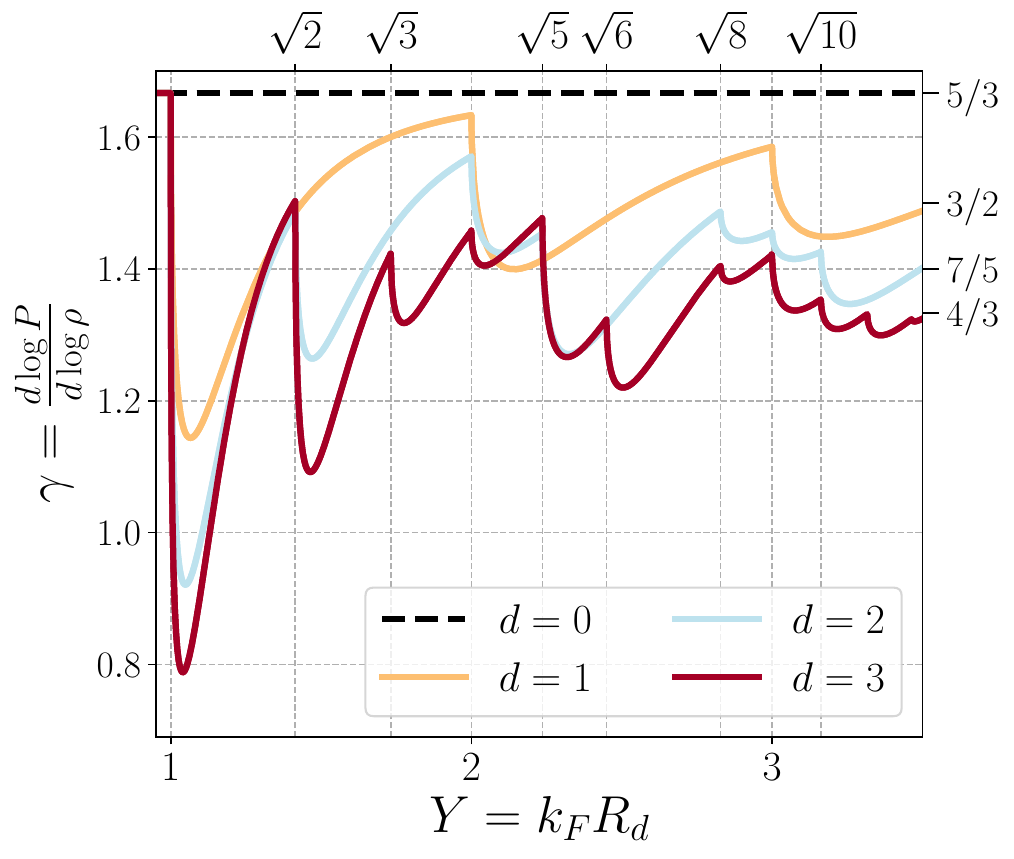}     \includegraphics[width=0.47\linewidth]{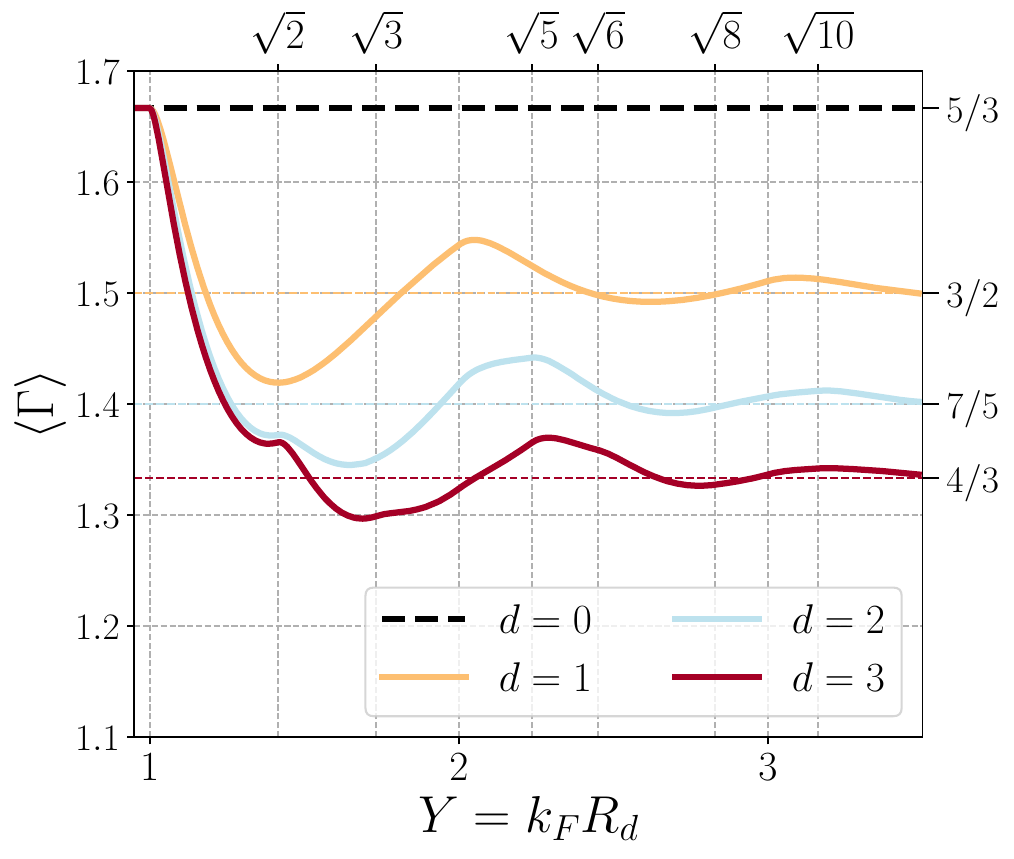}
   \caption{The polytropic index as a function of $Y = k_F R_d$ is shown in the left panel. The pressure averaged polytropic index is shown in the right panel.}
    \label{fig:gamma_Y}
\end{figure}

For higher Fermi momentum/densities, the effect of the extra-dimension(s) can significantly modify the thermodynamics of the system. In the left panel of figure~\ref{fig:gamma_Y}, we plot the polytropic index as a function of $Y$, whereas in the main text Fig.~\ref{fig:polytrop_d} displays $\gamma$ as a function of $\rho$. For clarity, we discuss the cases of one and two extra-dimensions. Situations with $d\geq3$ follow the same logic.

For $d=1$, as $k_F$ becomes an integer multiple of $1/R_d$, a new KK mode enters the sum of Eq.~(\ref{eq:Sk}) and a drastic softening of the equation of state is observed at all integer values of $Y=1, 2, 3,\ldots$~.\\

For $d=2$, due to the toroidal compactification of the extra-dimensions, the KK modes have multiplicity, and are labeled by two integers $\vec{n}=\left(n_1, n_2\right)$. Therefore, at $Y=1$, two different KK modes, namely $\left(1,0\right)$ and $\left(0,1\right)$ contribute, and a steeper softening of the EoS is observed. The next term in the KK sum is at $\vec{n}=\left(1,1\right)$, and the corresponding kink at $Y= \sqrt{1^2 + 1^2}=\sqrt{2}$ can be observed. Similar argument holds for larger multiplicities (or higher dimensions).\\

As the system becomes denser, the Fermi momentum grows and $Y\gg 1$. Once $k_F\gg 1/R_d$, few momentum modes are stricly 3-dimensional, and the system now behaves as a polytropic fluid of polytropc index $\frac{5+d}{3+d}$. This can be shown by taking the continuum limit of the discrete KK sums,  $S_\kappa \sim Y^d$. Finally, when graphed against $\rho$ instead of $k_F$, the period of the oscillations in $\gamma$ is stretched over a wide range, as $\rho\propto k_F^{3+d}$.\\

In the right panel of the figure we present the pressure averaged polytropic index, 
\begin{align}
    \left\langle \Gamma\right \rangle = \frac{\int dV P \times \gamma}{\int dV P}~.
\end{align}
As this is a volume integral over the polytropic index $\gamma$, the wiggles are inherited for the same values of $Y$.

\section{Approximate EoS and Chandrasekhar criterion}\label{app:LM_sols}

 Particles begin to delocalize in extra-dimensions when the condition $E_F \gtrsim E^{(1)}$ is satisfied. To understand the impact of extra-dimensions on the stability of self-gravitating systems we first construct an effective equation of state (EoS). As a zeroth order approximation, we consider a polytropic equation of state for non-relativistic DM fermions in the zero temperature limit of the form $P = K_d \rho^{\gamma_d}$. The constants $K_d$ and the index $\gamma_d = 1 + 2/(3+d)$ depend the number of spatial extra dimensions ($d$). For $3+d$ large spatial dimensions, the number density, energy density and pressure are given by
 \begin{align}
 n_{3+d}&=g\int\limits_0^{k_F} \frac{d^{3+d}k}{\left(2\pi\right)^{3+d}}=g\frac{\Omega}{\left(2\pi\right)^{3+d}}\frac{k_F^{3+d}}{3+d}\\
 \rho_{3+d}&=m n_{3+d}\\
 P_{3+d}&=g\int\limits_0^{k_F} \frac{d^{3+d}k}{\left(2\pi\right)^{3+d}}\frac{k^2}{\left(3+d\right)m}=g\frac{\Omega}{\left(3+d\right)\left(5+d\right)\left(2\pi\right)^{3+d}}\frac{k_F^{5+d}}{m}
 \end{align}
 where $\Omega=\frac{2\pi^{\frac{3+d}{2}}}{\Gamma\left(\frac{3+d}{2}\right)}$ is the value of the solid angle. From this we deduce the EoS of the $3+d$-dimensional non-relativistic free Fermi gas
 \begin{align}
P_{3+d}=K_d\left(\rho^{}_{3+d}\right)^{\gamma_d}
 \end{align}
 where
 \begin{align}
 \gamma_d&=\frac{5+d}{3+d}\\
K_{d}&=\frac{2\pi}{g^{\frac{2}{3+d}}}\frac{\left(\Gamma\left(\frac{5+d}{2}\right)\right)^{\gamma_d}}{\Gamma\left(\frac{7+d}{2}\right)}\frac{1}{m^{\gamma_{d}+1}}
 \end{align}

 Using dimensional arguments we construct an EoS that is effectively expressed in terms of 3D pressure and energy density. Assuming uniform particle distribution within the higher dimensional volume we write,  $\rho_{3+d} = \rho_{3}/V_{\rm extra}$ and $P_{3+d} = P_{3}/V_{\rm extra}$. The volume factor is denoted by $V_{\rm extra} = (2\pi R)^d$.  We then have the following piece-wise EoS,
\begin{eqnarray}\label{eq:eos_piece}
  P_{3}    &=& K_0\, \rho^{5/3}_{3}\quad \quad \quad \quad \quad \quad \quad \quad \quad \quad \quad \textrm{for } \rho \lesssim \rho_\star~, \\
  \bar{P}_{3}    &=& K_{d}\,\rho^{\gamma_d
  }_{3} \times V^{1-\gamma_d}_{\rm extra} \quad \quad \quad \quad \quad \quad \quad \textrm{for } \rho \gtrsim \rho_\star~.
\end{eqnarray}
From these expressions, it becomes clear that the equation of state becomes softer above the critical density $\rho_\star$ at which extra dimensions become important. Note that we have assumed that extra-dimensional phase space is uniformly occupied above $\rho_\star$.  From the above we construct a phenomenological equation of state that smoothly interpolates between the two cases,

\begin{eqnarray}\label{eq:eos_intp}
  P(\rho) &=&(1- f(\rho)) P_{3} + f(\rho) \bar{P}_{3}  ~,\\
  f(\rho) &=& \frac{1}{2} \left(1 + \tanh\left(\frac{\rho - \rho_\star}{\Delta} \right) \right)~.
  \end{eqnarray}
The variable $\Delta \ll \rho_\star$, controls the sharpness of excursions to extra-dimensions.\\

\noindent{\bf Mass-Radius relation:} The maximum mass is estimated by solving the Lane-Emden equation. At densities $\rho \gg \rho_\star$ the dominant contribution to the pressure comes from the $\bar{P}_3$ term.  The mass and radius can then be written as 

\begin{eqnarray}\label{eq:mass}
    M &=& 36 \pi  \left(\frac{(n+1)K_{d}\, V^{-1/n}_{\rm extra} }{4\pi G}\right)^{3/2} \rho_{\rm avg}^{\frac{3-n}{2 n}} \left(\frac{\xi_1}{3} \right)^{\frac{3(n+1)}{2 n}}  \left(-\theta^\prime(\xi_1)\right)^{\frac{3(n-1)}{2n}}~,\\
    R &=&  \left(\frac{(n+1)K_{d}\, V^{-1/n}_{\rm extra} }{4\pi G}\right)^{1/2} \rho_{\rm avg}^{\frac{1-n}{2 n}} \left(\frac{\xi_1}{3} \right)^{\frac{(1-n)}{2 n}} \xi_1 \left(-\theta^\prime(\xi_1)\right)^{\frac{(n-1)}{2n}} \label{eq:rads}
\end{eqnarray}
with $n=1/(\gamma-1)=(3+d)/2$, and the average density is denoted by $\rho_{\rm avg}$. We note values of the Lane-Emden parameters, and the corresponding mass in table~(\ref{tab:xi}).\\

For the case of $d=3$, the enclosed mass is independent of density, indicating the existence of critical maximum mass, i.e. equivalent of Chandrasekhar limit. Then the critical mass is 
\begin{equation}
    M_{\rm crit} =  10^{40}\, {\rm GeV}\left(\frac{\rm fm}{R_\star} \right)^{3/2} \left(\frac{5\,{\rm TeV}}{m} \right)^{7/2}~.
\end{equation}
For $n>3$ or $d>3$, the mass of the configuration decreases as the central density is increased, possibly signaling instability. Plugging in the average density to be $\rho_{\rm avg} = m/(3\pi^2 R^3_\star)$, we get the same maximum critical mass noted above. An important feature of extra-dimensional scenario is that the DM cloud is destabilized when the DM particles are non-relativistic, $\rho_{\text{crit}}\ll m_{\text{DM}}^4$. In the limit $R_\star\rightarrow 0$, the dark matter cloud can be destabilized only by relativistic effects, akin to the gravitational collapse of heavy neutron stars. The collapsing mass corresponds to the standard Chandrasekhar limit which is realized when $M_{\text{Ch}}<M_{\text{crit}}$. The Chandrasekhar mass is given by $M_{\text{Ch}}\approx M_{\text{Pl}}^3/m_{\text{DM}}^2$.

\begin{table}[htb]
\centering
\begin{tabular}{ccccc}
\toprule
$d$ & $n$ & $\xi_1$ & $-\theta^\prime(\xi_1)$ & $M/ (M_{\rm pl}/m)^3$\\
\midrule
0 & 3/2 & 3.653 & 0.2033 & $\approx 0.05 \,\rho_{\rm avg}^{1/2}/m$\\
1 & 2 & 4.353 & 0.1273 & $\approx 0.01 \,\rho_{\rm avg}^{1/4} \left(\frac{1}{R_\star m}\right)^{3/4}$\\
2 & 5/2 & 5.355 & 0.0763 & $\approx 10^{-3} \,\rho_{\rm avg}^{1/10} m^{6/10} \left(\frac{1}{R_\star m}\right)^{6/5}$\\
3 & 3 & 6.897 & 0.0424 & $\approx 10^{-3} \left(\frac{1}{R_\star m}\right)^{3/2} m$\\
4 & 7/2 & 9.536 & 0.0208 & $\approx 10^{-3} \,\rho_{\rm avg}^{-1/14} m^{9/7} \left(\frac{1}{R_\star m}\right)^{12/7}$\\
\bottomrule
\end{tabular}
\caption{\label{tab:xi}
Values of $\xi_1$ and $\theta^\prime(\xi_1)$ for selected $(n,d)$.}
\end{table}

\end{document}